%% file: corr_energy_arxiv_v4.tex
\documentclass[11pt,accepted=2017-10-06,onecolumn,a4paper]{quantumarticle}

\pdfoutput=1

\usepackage{amsmath,amssymb,verbatim,graphicx,color}
\usepackage{authblk}
\usepackage[labelformat=simple]{subcaption}
\usepackage{quant_defs}
\usepackage[retainorgcmds]{IEEEtrantools}
\usepackage[square,comma,numbers,sort&compress,merge]{natbib}
\usepackage[bookmarks=false]{hyperref}
\makeatletter
\fancy@setoffs

\makeatletter
\renewcommand\p@subfigure{\thefigure}
\makeatother

\newcommand{\vect}[1]{\boldsymbol{#1}}
\newcommand{\rz}{\mathrm{z}}
\newcommand{\rg}{\mathrm{g}}

\newcommand{\setQ}{\mathcal{Q}}
\newcommand{\setQa}{\overline{\mathcal{Q}}}
\newcommand{\setQp}{\widehat{\mathcal{Q}}}
\newcommand{\setCa}{\overline{\mathcal{C}}}
\newcommand{\setCp}{\widehat{\mathcal{C}}}
\newcommand{\setDa}{\overline{\mathcal{D}}}
\newcommand{\setDp}{\widehat{\mathcal{D}}}
\newcommand{\rx}{\mathrm{x}}
\newcommand{\h}{\omega}

\DeclareMathOperator{\erf}{erf}

\include{jnls}
\newcommand{\emailto}[1]{\thanks{\href{mailto:#1}{#1}}}

\begin{document}
\title{Semi-device-independent framework\\
  based on natural physical assumptions}
\author[1,2]{Thomas~Van~Himbeeck}
\emailto{thomas.van.himbeeck@ulb.ac.be}
\author[3]{Erik~Woodhead}
\author[2]{Nicolas~J.~Cerf}
\author[2]{Ra\'{u}l~Garc\'{i}a-Patr\'{o}n}
\author[1]{\mbox{Stefano~Pironio}}
\affil[1]{Laboratoire d'Information Quantique,
  Universit\'{e}~libre~de~Bruxelles~(ULB), Belgium}
\affil[2]{Centre for Quantum Information and Communication,
  Universit\'{e}~libre~de~Bruxelles~(ULB), Belgium}
\affil[3]{ICFO -- Institut~de~Ci\`e{}ncies~Fot\`o{}niques, The Barcelona
 Institute of Science and~Technology,\\
 08860 Castelldefels (Barcelona), Spain}


\date{October 6, 2017}

\maketitle
\begin{abstract}
  The semi-device-independent approach provides a framework for
  prepare-and-measure quantum protocols using devices whose behavior does not
  need to be characterized or trusted, except for a single assumption on the
  dimension of the Hilbert space characterizing the quantum carriers. Here,
  we propose instead to constrain the quantum carriers through a bound on the
  mean value of a well-chosen observable. This modified assumption is
  physically better motivated than a dimension bound and closer to the
  description of actual experiments. In particular, we consider quantum
  optical schemes where the source emits quantum states described in an
  infinite-dimensional Fock space and model our assumption as an upper bound
  on the average photon number in the emitted states. We characterize the set
  of correlations that may be exhibited in the simplest possible scenario
  compatible with our new framework, based on two energy-constrained state
  preparations and a two-outcome measurement. Interestingly, we uncover the
  existence of quantum correlations exceeding the set of classical
  correlations that can be produced by devices behaving in a purely
  pre-determined fashion (possibly including shared randomness). This feature
  suggests immediate applications to certified randomness generation. Along
  this line, we analyze the achievable correlations in several
  prepare-and-measure optical schemes with a mean photon-number constraint
  and demonstrate that they allow for the generation of certified
  randomness. Our simplest optical scheme works by the on-off keying of an
  attenuated laser source followed by photocounting. It opens the path to
  more sophisticated energy-constrained semi-device-independent quantum
  cryptography protocols, such as quantum key distribution.
\end{abstract}

\section{Introduction}

Understanding the nature and extent of correlations that distinct systems may
display is a central problem in many applications of quantum physics. In
particular, it is an essential stone for developing device-independent (DI)
quantum information protocols
\cite{ref:my1998,ref:bhk2005,ref:c2006,*ref:ck2011,ref:ab2007,ref:ruv2013,
  ref:ms2014,ref:afrv2016}, where the correlations that are observed between
separate quantum devices provide a guarantee that the protocol performs as
expected. This guarantee follows independently of any assumptions on the
local behavior of the quantum devices, hence its name, but it must
necessarily rely on some specific constraints on the information that they
exchange. Indeed, if arbitrary, unlimited communication is allowed between
the devices, any kind of correlations can be generated, even in a scenario
restricted to classical physics.

In the standard DI framework based on Bell non-locality \cite{ref:bc2014},
the constraint on communication is maximal: the separate devices are not
allowed to communicate any type of information, neither classical nor
quantum. This no-communication constraint has the conceptual advantage of
having a clear physical and operational significance. In particular, it is in
principle possible to enforce it without knowledge of the internal behavior
of the devices, i.e., by adequate shielding or space-like separation of the
devices. However, the generation of useful, non-classical correlations in the
absence of quantum communication must then necessarily rely on
(loophole-free) entanglement, which presently represents a serious obstacle
to practical DI applications.

This difficulty has motivated the development of an alternative framework for
DI applications, which is inspired by the traditional prepare-and-measure
implementation of quantum key distribution and where communication is allowed
between the quantum devices. As noted above, a constraint, though, must be
put on this communication and it is usually formulated as a bound on the
Hilbert space dimension of the exchanged quantum messages
\cite{ref:bp2008,ref:gb2010,ref:bq2014}. With such a constraint, useful
non-classical correlations can already be generated by restricting the
communication to qubits or qutrits in a purely prepare-and-measure scenario,
without the need of entanglement, which provides a clear advantage from the
implementation point of view. Several protocols for randomness generation
(RNG) \cite{ref:lb2015} and quantum key distribution (QKD)
\cite{ref:pb2011,ref:wp2015} have been introduced within this framework,
which is usually referred to as ``semi-device-independent'' (semi-DI). The
downside, however, is that the dimension assumption, even if it represents a
convenient abstraction for a theorist, is only an idealization. Carriers of
quantum information, such as photons, live in an infinite Hilbert space, and
assuming that information is encoded in only a few degrees of freedom is not
justified without some intricate characterization of the devices (hence the
terminology ``semi device independent'').

In this paper, we propose a physically better motivated approach for
constraining the exchanged quantum messages in a semi-DI framework. We
express the restriction on the exchanged states in terms of the mean values
of some well-chosen observable, such as the energy. As a simple example,
consider the case where the Hilbert space carrying the quantum messages is
the Fock space of several quantum optical modes. This is the appropriate
space to describe quantum optics experiments, including those demonstrating
results based on dimension bounds, in which attenuated laser sources
\cite{ref:abc2012} or non-ideal heralded photon sources
\cite{ref:hgm2012,ref:lb2015} are used. In this context, the emitted states
can in principle occupy an infinite-dimensional space so that, instead of
putting a limit on the dimension, it is much more natural to constrain the
average number of photons. The corresponding observable would then be the
photon-number operator, which has a clear physical significance. Alternatively, we could constrain the energy contained in one or more frequency modes containing the quantum message, as the two are closely related. This is thus a natural substitute for the dimension of a
finite Hilbert space. 
Moreover, designing devices in such a way that the
average photon number does not exceed a given threshold or verifying
experimentally that it does not exceed such a threshold will typically
require a less detailed modeling of the devices than would be needed to
verify, e.g., that the emitted states span a Hilbert space of a given
dimension.

A prerequisite for the development of any DI or semi-DI protocol is to
examine the set of correlations that are available under the assumptions
considered. Much work has been done specifically on this question in the
standard settings based on non-locality and dimension bounds, see e.g.\
\cite{ref:t1987,ref:ww2001,ref:m2003,ref:c2004,ref:npa2007,*ref:npa2008,ref:dt2008,ref:bc2014}
and \cite{ref:bp2008,ref:gb2010,ref:bq2014,ref:bnv2013,ref:nv2015}. Here, we
fully characterize analytically the set of available correlations in the
simplest scenario compatible with our general framework. This uncovers
interesting features suggesting immediate applications to randomness
generation, which will be fully developed in a forthcoming publication
\cite{ref:vhPREP}.

In Section~\ref{sec:frame}, we introduce the general framework of semi-DI
prepare-and-measure scenarios and modify it to account for a physical
constraint (mean value of an observable) instead of a dimension bound. We
define a simple setting with two state preparations and a single measurement
with binary outcomes, which suffices to produce a separation between the
quantum and classical correlations, and subsequently suggest a few simple
potential implementations using currently accessible quantum optics
technology. The quantum region (i.e., the set of available quantum
correlations) is analyzed in Sections~\ref{sec:Q} and \ref{sec:genq}, while
the classical region (i.e., the set of available correlations arising from a
mixture of classical deterministic behaviors) is studied in
Section~\ref{sec:c}. Then, in Section~\ref{sec:d}, with an eye toward the
application to certified random number generation, we characterize an
intermediate deterministic region; correlations outside of this region are
those for which randomness can be certified for a specified input. We also show
that correlations outside this intermediate region are achievable with simple
optical implementations. Finally, we conclude in Section~\ref{sec:conclusion}
and discuss other possible applications.

\section{Semi-DI setting with a physical constraint}
\label{sec:frame}

\subsection{Definition of the general model}

\begin{figure}[tbp]
  \centering
  \includegraphics{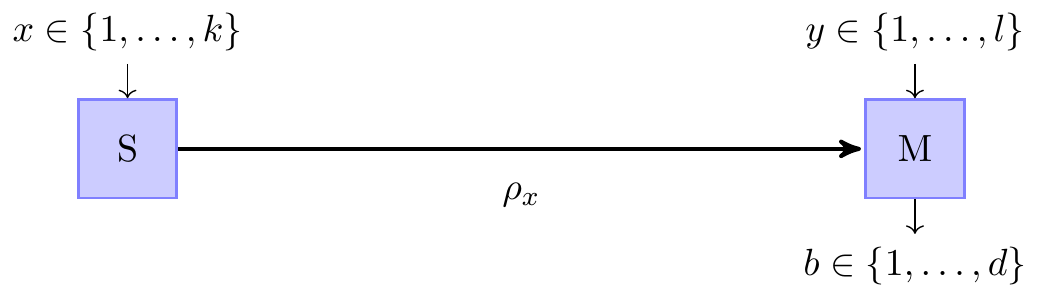}
  \caption{A general prepare-and-measure scenario. Source (S) emits one of
    $k$ states $\rho_{x}$ depending on an input $x \in \{1, \dotsc, k\}$. A
    measurement device (M) performs one of $l$ measurements on the state
    received, depending on an input $y \in\{1, \dotsc, l\}$, and registers an
    outcome $b \in \{1, \dotsc, d\}$. The behaviors of S and M are not
    characterized and could even depend on shared hidden parameters
    $\lambda$. But we trust that the prepared states satisfy constraints that
    are expressed in term of the expectations $\Tr[H \rho_{x}]$ of some given
    Hermitian operator $H$.}
  \label{fig:pm_general}
\end{figure}

Let us first remind the general framework of semi-DI prepare-and-measure
scenarios. As depicted in Fig.~\ref{fig:pm_general}, a source S is linked
through a quantum channel to a measurement device M\@. On the source S, an
input $x \in \{1, \dotsc, k\}$ can be selected, resulting in the emission of
an unknown quantum state $\rho_x$. The state is then measured by M, according
to a measurement selected through an input $y \in \{1, \dotsc, l\}$, and
yields an outcome $b \in \{1, \dotsc, d\}$. This later process is
characterized by a set of unknown measurement operators $\{M_{b|y}\}$.

To an external observer that has access only to the inputs and output of S
and M, the joint behavior of the two devices is completely characterized by
the probabilities
\begin{equation}
  \label{eq:pbyx}
  P(b|y,x) = \Tr[M_{b|y} \, \rho_x] \,.
\end{equation}
More generally, the behavior of the two devices could be correlated through
dependence on an additional hidden random parameter $\lambda$ shared between
the devices, in which case the probabilities take the
more general form
\begin{equation}
  \label{eq:pbyxl}
  P(b|y,x)
  = \sum_\lambda p_\lambda \Tr\bigsq{M^\lambda_{b|y} \, \rho^\lambda_x} \,.
\end{equation}

In the semi-DI approach, no detailed assumptions are made on the states and
measurements underlying the correlations $P(b|y,x)$, except for a specific
constraint on the messages $\rho_x$. Here, we propose to express such a
constraint in terms of an observable $H$, describing a physical property of
the emitted states $\rho_x$ that we trust or on which we have control (more
generally, one could introduce several such observables). A restriction on
the quantum messages $\rho_x$ can then be formulated as a constraint on the
corresponding mean values $H_x = \Tr[H \rho_x]$ of this observable.

Note that contrarily to the states $\rho_x$ and measurement operators
$M_{b|y}$, which are a priori unspecified and unknown, the observable $H$
must be well defined. It is also implicit that it should be defined on some
given (possibly infinite) Hilbert space $\mathcal{H}$ describing the physics
of the quantum systems emitted by S\@. Thus, one should also assume that
$\rho_x$ is defined on $\mathcal{H}$.

This general formulation encompasses the usual dimension assumption, for
instance, by defining $H$ as the projector onto a qudit subspace of
$\mathcal{H}$ and requiring $H_x = 1$ for all $x$. Expressing the dimension
assumption in this form has the merit of making explicit that the message
qudits live in a subspace of a larger Hilbert space $\mathcal{H}$, which in
practice must also be properly defined and characterized if one wants to make
sure that information is propagating in the relevant subspace and not in
possible additional degrees of freedom, which in a cryptographic protocol
could be exploited by an eavesdropper (side channels).

As stated in the introduction, the main interest of our more general
formulation, however, is that it can be used to model communication
constraints that are more natural and better motivated physically than the
usual dimension bounds. A particular example is the case where $H$ is the
photon-number operator of a quantum optical system.

With this application in mind, we will consider in this work two types of
constraints on the mean values of $H$. The first, which we denote the
\emph{max-average assumption}, corresponds to assuming upper bounds
\begin{equation}\label{eq:bounds}
  H_x = \Tr[H \rho_{x}]
  = \sum_\lambda p_\lambda \Tr[H \rho^\lambda_x] \leq \h_x \,,
  \quad \forall x
\end{equation}
on the mean values $H_x$ for given thresholds $\h_x$. For instance, if $H$ is the
photon-number operator, we may trust that for all states $\rho_x$ emitted by
the source, the mean photon numbers $H_x$ are below some threshold, though we
may not know what the actual photon number is.

In the case where the states emitted by S vary from run to run according to
some random parameter $\lambda$, the max-average assumption only bounds the
mean value $H_x$ averaged over all possible values of $\lambda$. But it does
not constrain the maximum values of $H_{x|\lambda}=\Tr[H \rho^\lambda_x ]$,
which could in principle be arbitrarily high. It is therefore natural to
introduce another (stronger) assumption, which we call the \emph{max-peak
  assumption}, according to which
\begin{equation}
  \label{eq:bounds2}
  \max_\lambda H_{x|\lambda}
  = \max_\lambda \Tr[H \rho^\lambda_x] \leq \h_x \,,
  \quad \forall x \,.
\end{equation}
Note that if $H$ is, e.g., the photon-number operator, this second
condition still allows for fluctuations in the photon number within
each state $\rho^\lambda_x$ and does not correspond to a truncation of
the Fock space, as the constraint only imposes a bound on the
\emph{mean} values $\Tr[H \rho^\lambda_x] $ of $H$ for every
$\rho^\lambda_x$. In particular, the states $\rho^\lambda_x$ could
have a non-zero amplitude in any of the number-basis states.

These two possible physically-motivated constraints on the communicated
quantum messages will be analyzed on specific examples later on. The
max-average assumption has the advantage that it could in principle be
verified ``from the outside'' by performing tests on the average emitted
states $\rho_x$ without any knowledge of the internal behavior of the
source. Verifying that the max-peak assumption is satisfied, on the other
hand, would typically rely on some modeling of the source. Its main advantage
is that it is more constraining and thus can certify useful properties that
wouldn't be certified using the average-peak assumption (see examples in
Sections~\ref{sec:c} and \ref{sec:d}).

\subsection{The simplest setting: two inputs and two outputs}

\begin{figure}[tbp]
  \centering
  \includegraphics{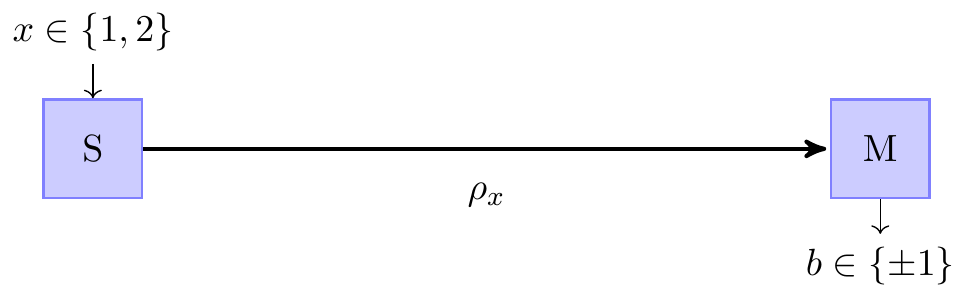}
  \caption{The prepare-and-measure scenario considered here, a special case
    of Fig.~\ref{fig:pm_general}.}
  \label{fig:pm_setup}
\end{figure}

In the rest of this paper, we consider the simple situation where the source
S has two possible inputs $x\in\{1,2\}$ and the measurement device M has no
input (i.e., $y\in \{1\}$) and two possible outputs, which we denote
$b\in\{\pm 1\}$ for convenience, see Fig.~\ref{fig:pm_setup}. Note that this
corresponds to the simplest non-trivial prepare-and-measure scenario. Indeed,
M must obviously output at least two different possible values, and S must
have at least two different preparation choices, otherwise no quantum
messages are needed and any observed behavior can be classically simulated by
S and M. 

This prepare-and-measure scenario is simpler than any possible scenario based
on a dimension bound, for which one must have at least three choices on the
source, i.e., $x \in \{1, 2, 3\}$, and any measurement device outputting a
single bit $b \in \{\pm 1\}$ should have at least two inputs, i.e.,
$y \in \{1, 2\}$.  Indeed, since the smallest dimension bound corresponds to
one qubit, the channel connecting S to M always has a capacity of at least 1
bit under a dimension assumption. This implies that the number of inputs on S
should be larger than two, because otherwise the input $x$ can be encoded
perfectly in the channel and transmitted to M, who knowing $x$ can now
generate an output $b$ compatible with any probability distribution
$p(b|x,y)$. There should also be a number of binary measurements greater than
one, because otherwise S could locally choose a value $b\in\{\pm 1\}$
compatible with any probability distribution $p(b|x)$ and simply send that
value $b$ to M through the channel. Strikingly, such strategies are not
available under the assumptions that we consider here, since, as we will see,
they constrain the classical channel capacity to be sub-unity by forcing the
two emitted states $\rho_1$ and $\rho_2$ to have some non-zero overlap.

In our scenario, the joint behavior of S and M is thus characterized by the
four probabilities $P(b|x) = \Tr[M_{b} \, \rho_x]$ where $x = 1, 2$ and
$b = \pm 1$. It will be convenient to work with the equivalent representation
\begin{equation}\label{eq:ex}
  E_x = \Tr[M \rho_x] \qquad (x=1,2) \,,
\end{equation}
where $E_x = P(+1|x) - P(-1|x)$ is the expectation value of the observable
$M = M_{+1} - M_{-1}$, with $- \id \leq M \leq \id$. The value of $E_x$
characterizes the bias of the output $b$ toward $+1$ or $-1$ for a given
input $x$. We refer in the following to the quantities $E_x$ as
``correlations'' as they represent how the output of M is correlated to the
input of S\@. For instance if $E_1 = 1$ and $E_2 = -1$, the output of M is
completely correlated to the input of S\@. More precisely, the presence of
correlations is actually reflected by the fact that
$\abs{E_-} = \abs{E_1 - E_2} > 0$. In particular, a value $\abs{E_-} > 0$
implies that the measurement device M can (at least partly) distinguish the
two states $\rho_1$ and $\rho_2$. We will see further below that the quantity
$E_-$ plays a special role in our analysis, analogous to a Bell expression in
the context of non-locality.

Having defined the general behavior of S and M, we now specify the properties
of the observable $H$ that we use to model our assumptions on the quantum
messages. As noted above, in our scenario the two states $\rho_1$, $\rho_2$
should have a non-zero overlap, otherwise they could encode faithfully the
two values $x = 1, 2$ and our entire problem would become trivial. One simple
possibility for satisfying this condition in an optical system is simply to
have both states $\rho_1$, $\rho_2$ sufficiently close to the vacuum state
$\ket{0}$. In a multimode system with a discrete and finite number of
possible mode frequencies $\omega$, this amounts to upper bounding the
expectation values of the photon-number operator
$H = \sum_{\omega} a_\omega^{\dagger} a_\omega$ or the energy operator
$H = \sum_{\omega} \hbar \omega \, a_\omega^{\dagger}
a_\omega$. Alternatively, one could directly bound the weight of the
non-vacuum component, i.e., the expected value of the non-vacuum projector
$H = \id - \proj{0}$. More generally, the condition that $\rho_1$, $\rho_2$
have a non-zero overlap is satisfied if they are both close to some given
reference state $\proj{\phi}$, i.e., if the expectation values $H_x$ of the
observable $H = \id - \proj{\phi}$ are below some sufficiently small
thresholds.

Formally, all the above examples correspond to constraints on the expected
values of an observable H satisfying the two following conditions:
\begin{enumerate}
  \item H has a non-degenerate ground state,
  \item H has a finite gap.
\end{enumerate}
The results that we will derive below apply to any observable H satisfying
these two conditions, independently of their physical meaning. Without loss
of generality, we can assume (if necessary by rescaling $H$) that the ground
state eigenvalue is 0 and the gap is~1. In the following, we let $|0\rangle$
denote the ground state of $H$.

Before characterizing, in Sections~\ref{sec:Q} and \ref{sec:genq}, the set of
possible correlations $E_x$ between S and M in terms of the constrained mean
values $H_x$ of such observables $H$, we briefly describe for concreteness
some standard optical circuit implementations that can be analyzed in our
framework.

\subsection{Examples of optical circuits}
\label{sec:impl}

\begin{figure}[tbp]
  \begin{subfigure}{\linewidth}
    \centering
    \includegraphics{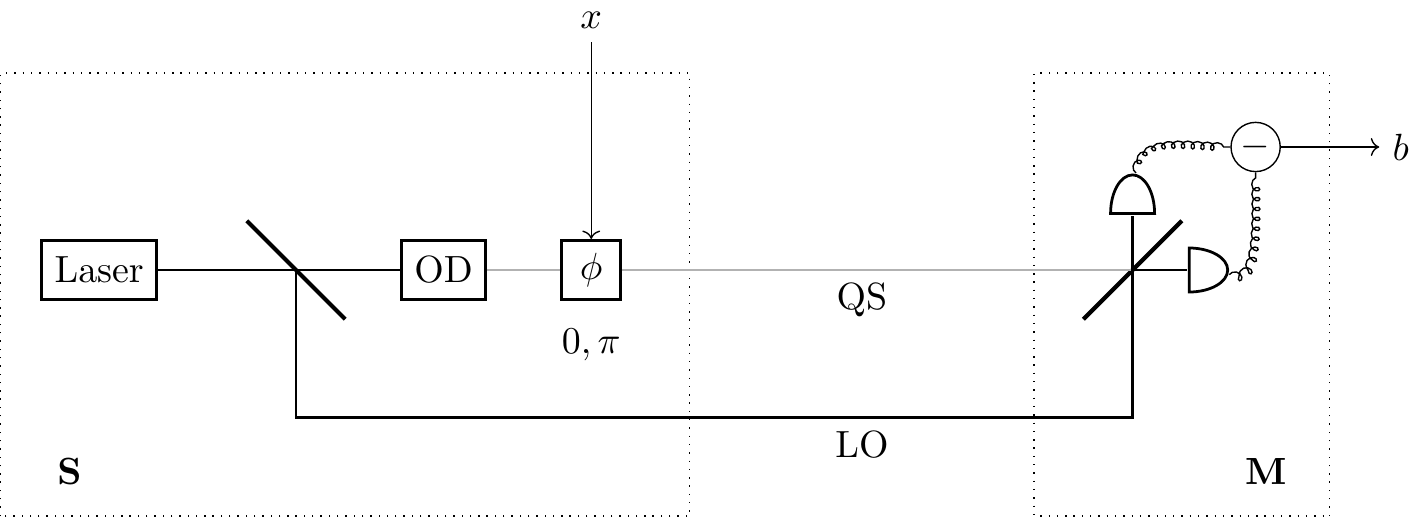}
    \caption{BPSK implementation. The source (S) consists of a laser that
      produces a coherent pulse which is sent through an unbalanced beam
      splitter. The intense reflected beam is sent to the measurement device
      (M) as a phase reference, i.e., local oscillator (LO); the transmitted
      beam is highly attenuated at an optical density (OD) and phase shifted
      by $0$ or $\pi$ depending on $x$ at a phase shifter. This is the
      quantum signal (QS) sent to M\@. M then performs a homodyne measurement
      on QS: the two beams interfere on a balanced beam splitter and two
      photodiodes measure the intensities of the resulting beams. The
      difference of the two intensities is proportional to the quadrature $X$
      of the quantum signal. Finally, M outputs the sign of $X$. Note that
      the LO does not depend on the input $x$ and can be modeled as shared
      randomness.}
    \label{fig:impl_bpsk}
  \end{subfigure}

  \vspace{0.8cm}

  \begin{subfigure}{\linewidth}
    \centering
    \includegraphics{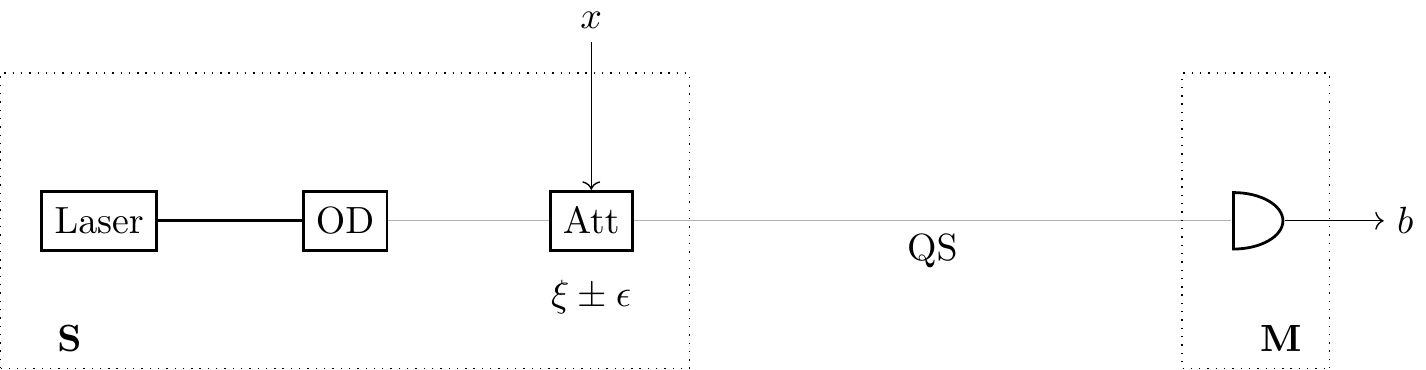}
    \caption{2ASK implementation. The source (S) consists of a laser that
      produces a coherent pulse, which is, at first, highly attenuated at an
      optical density (OD), and then sent through a variable attenuator
      (Att), so that, depending on $x$, the amplitude of the resulting
      coherent states is $\xi \pm \epsilon$. This is the quantum signal (QS)
      sent to the measurement device (M). M is a single-photon detector that
      outputs $b = +1$ if it clicks, and $b = -1$ otherwise.}
    \label{fig:impl_2ask}
  \end{subfigure}

  \vspace{0.8cm}

  \begin{subfigure}{\linewidth}
    \centering
    \includegraphics{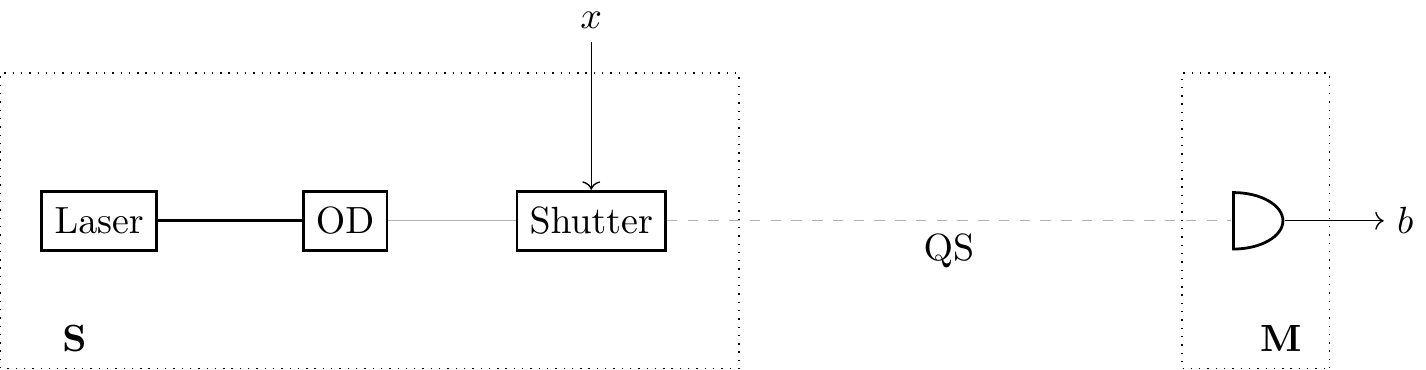}
    \caption{OOK implementation. The source (S) consists of a laser that
      produces a coherent pulse, which is, at first, highly attenuated at an
      optical density (OD). A controllable shutter then transmits or blocks
      the beam depending on the value of $x$. The measurement device is a
      single-photon detector that outputs $b = +1$ if it clicks, and $b = -1$
      otherwise. Alternatively to the use of a shutter, the laser can simply
      be turned on and off depending on $x$.}
    \label{fig:impl_ook}
  \end{subfigure}
  \caption{Three experimental implementation propositions.}
\end{figure}

\subsubsection{Binary Phase-Shift Keying (BPSK)}
\label{sec:impl_bpsk}

Our first optical implementation corresponds to the Binary Phase-Shift Keying
(BPSK) scheme and is illustrated in Fig.~\ref{fig:impl_bpsk}. We consider a
single optical mode described over the phase space $(X,P)$ of the two
quadratures of light ($[X,P]=\frac{i}{2}$). Depending on $x$, the source
prepares one of two coherent states $\ket{\phi_1} =\ket{\xi}$ or
$\ket{\phi_2} =\ket{-\xi}$, where $\xi$ is a small positive real parameter,
so that both states are close to the vacuum $\ket{0}$. Although these two
states have a non-zero overlap $e^{-2\abs{\xi}^2}$, it is possible to partly
distinguish them by performing a homodyne measurement of the quadrature
$X$. In particular if we define the output of the measurement device M as
$b=\text{sign}(X)$, then a straightforward calculation gives
\begin{IEEEeqnarray}{c+c}
  \label{eq:corr1}
  E_1 = \erf(\sqrt{2} \xi) \,, & E_2 = -\erf(\sqrt{2} \xi)\,,
\end{IEEEeqnarray}
i.e., we observe a correlation between the sign of $b$ and the input $x$
whose strength depends on the value $\xi$.

A semi-DI analysis for this setup (see Sections~\ref{sec:c} and \ref{sec:d})
is possible based only on the assumption that the source emits optical
systems with low mean photon number (choosing
$H = \sum_{\omega} a_\omega^{\dagger} a_\omega$) or small non-vacuum
component (choosing $H = \id - \proj{0}$). The coherent states
$\ket{\pm \xi}$ have a mean photon number $\abs{\xi}^2$ and non-vacuum
component $1 - e^{-\abs{\xi}^2}$, where the mean photon number is larger than
the non-vacuum component: $\abs{\xi}^2 > 1 - e^{-\abs{\xi}^2}$. Thus for
given $\xi$, one can impose more constraining bounds if one defines $H$ as
the non-vacuum projector rather than the photon-number operator and we will
thus make the former choice in the following. The difference between
$1-e^{-\abs{\xi}^2}$ and $\abs{\xi}^2$, though, is negligible when $\xi$ is
small and this choice does not fundamentally affect our results. When
performing the semi-DI analysis of the above setup, we will thus take
$H=\id-\proj{0}$ and upper bound the expectation values of this operator
through the max-average assumption \eqref{eq:bounds} or the max-peak
assumption \eqref{eq:bounds2} using thresholds
\begin{equation}
  \label{eq:thresh1}
  \h_{1} = \h_{2} = 1 - e^{-\abs{\xi}^2} \,.
\end{equation}

Note that in the implementation illustrated in Fig.~\ref{fig:impl_bpsk}, in
addition to the quantum states $\ket{\phi_x}$, the source also emits an
intense reference laser beam that serves as a local oscillator to define the
phase of these states. This local oscillator can, however, be modeled as
shared randomness (for instance it could exit the source S before the input
$x$ is chosen). Alternatively, one could also consider more involved
implementations where a phase synchronization between S and M can be achieved
without the need for the transmission of a local oscillator signal between S
and M \cite{ref:ql2015}.

\subsubsection{2-level Amplitude-Shift Keying (2ASK)}
\label{sec:impl_2ask}

Our second example corresponds to the 2-level Amplitude-Shift Keying (2ASK)
scheme and is illustrated in Fig.~\ref{fig:impl_2ask}. The source emits the
two coherent states $\ket{\phi_1} =\ket{\xi + \epsilon}$ or
$\ket{\phi_2} = \ket{\xi - \epsilon}$, where $\xi$ and $\epsilon$ are two
real, positive parameters. The measurement device is a photodetector that
outputs $b = -1$ if no photon is detected and $b = +1$ if at least one photon
has been detected. This setup generates the correlations
\begin{IEEEeqnarray}{c+c}
  \label{eq:corr2}
  E_1 = 1 - 2 e^{-(\xi+\epsilon)^2} \,,
  & E_2 = 1 - 2 e^{-(\xi-\epsilon)^2} \,.
\end{IEEEeqnarray}
Suitable choices for $\xi$ and $\epsilon$ can easily be found so that
$\abs{E_1 - E_2} > 0$, i.e., such that the output of M and the input of S are
correlated.

In this second example, the two states $\ket{\phi_1} = \ket{\xi + \epsilon}$ and
$\ket{\phi_2} = \ket{\xi - \epsilon}$ are not necessarily close to the vacuum
if $\xi$ is large, but they are close to the intermediate state
$\ket{\xi}$. We thus define for the purpose of the semi-DI analysis of this
setup (see Sections~\ref{sec:c} and \ref{sec:d}) the observable $H$ as
$H = \id - \proj{\xi}$, which measures the proximity of the states
$\ket{\phi_x}$ to the reference state $\ket{\xi}$. Specifically, we have
$\bra{\phi_x} H \ket{\phi_x} = 1 - e^{-\epsilon^2} \simeq \epsilon^2$ for
$x = 1, 2$. Thus we will constrain the expected values of $H$ using
thresholds
\begin{equation}
  \label{eq:thresh2}
  \h_{1} = \h_{2} = 1 - e^{-\epsilon^2} \,.
\end{equation}
Note that, contrarily to the previous example, such an assumption on the
expectation values of $H$ may actually require a more detailed characterization of the source. First, the observable does not correspond to a natural property, like the energy, and, second, an experimental verification of the assumption would require a displacement operation with a beam that is coherent with the quantum message followed by a photon number measurement. It is thus probably not the most
natural in a semi-DI setting. We nevertheless include this example to stress
that our formulation can be adapted to different constraints on the source.

\subsubsection{On-Off Keying (OOK)}
\label{sec:impl_ook}

Our final example corresponds to the On-Off Keying (OOK) scheme and is
illustrated in Fig.~\ref{fig:impl_ook}. When $x = 1$, the source emits a
coherent state $\ket{\xi}$ where $\abs{\xi}^2$ is small, and when $x = 2$ it
emits the vacuum state $\ket{0}$. The measurement performed at M corresponds,
as in the previous example, to a photodetector that outputs $b = -1$ if no
photon is detected and $b = +1$ if at least one photon has been
detected. This yields
\begin{IEEEeqnarray}{c+c}
  \label{eq:corr3}
  E_1 = 1 - 2 e^{-\abs{\xi}^2} \simeq -1 + 2 \abs{\xi}^2 \,, &
  E_2 = -1 \,.
\end{IEEEeqnarray}
That is, when $x = 2$ one obviously observes the result $b = -1$ with
certainty, while when $x = 1$, there is a non-zero probability
$1 - e^{-\abs{\xi}^2} \simeq \abs{\xi}^2$ to obtain the outcome $b = +1$. The
measurement performed at M can thus be interpreted as a partial unambiguous
discrimination of the two states $\rho_1$, $\rho_2$ in the sense that when we
find $b = +1$, we are sure that the state sent was $\rho_1$, but we cannot
conclude anything definite when $b = -1$.

As in the first example, the semi-DI analysis for this setup (see
Sections~\ref{sec:c} and \ref{sec:d}) will only rely on the assumption that
the source emits optical systems with low non-vacuum component, i.e., we use
$H = \id - \proj{0}$. The non-vacuum component is $1 - e^{-\abs{\xi}^2}$ for
$x = 1$ and $0$ for $x = 2$ and we will thus use the thresholds
\begin{IEEEeqnarray}{c+c}
  \label{eq:thresh3}
  \h_1 = 1-e^{-\abs{\xi}^2} \,, & \h_2 = 0 \,.
\end{IEEEeqnarray}
Note that contrarily to the two above examples, here we bound differently the
expectation values of $H$ in the case $x=1$ and $x=2$ since the
implementation is not symmetric with respect to the two
situations. Alternatively to bounding the non-vacuum component, we could use
the mean photon number, but the difference would be negligible when $\xi$ is
small and making one choice or the other does not fundamentally affect our
results.

\section{Pure-state quantum correlations}
\label{sec:Q}

As we have explained earlier, a prerequisite for the development of any DI or
semi-DI protocol is to examine the set of correlations that are available
under the assumptions considered. One of our objectives is thus to
characterize the most general set of quantum correlations $(E_1, E_2)$
compatible with arbitrary implementations of our prepare-and-measure
scenario.

In general, the source S and the measurement device M could behave in a way
depending on shared random parameters $\lambda$. As a first step, it is
useful to consider the case where the devices do not exploit such shared
randomness and where in addition the source S emits \emph{pure} states
$\phi_1$, $\phi_2$. In Section~\ref{sec:genq}, we will relax these conditions
and consider completely general implementations.

Let us therefore define the set of pure-state correlations as
\begin{equation}
  \label{eq:Q}
  \setQ_{H_1, H_2} = \bigbr{(E_{1}, E_{2})
    \bigm| E_{x} = \Tr[M \phi_{x}] , \,
    H_{x} = \Tr[H \phi_{x}] \text{ for } x = 1, 2} \,,
\end{equation}
that is, the set of possible values $(E_1, E_2)$ that are attainable with
arbitrary pure states $\phi_x = \proj{\phi_{x}}$ and an arbitrary measurement
operator $M$ satisfying $- \id \leq M \leq \id$, and which are compatible
with given expectation values $(H_1, H_2)$ for an observable $H$ with
non-degenerate ground state, lowest eigenvalue 0, and finite gap 1. To
simplify the notation, we will often write $\setQ$ for $\setQ_{H_1, H_2}$
with the implicit understanding that the values $(H_1, H_2)$ are fixed.

A first observation is that if $H_1 + H_2 = 1$ then any correlations $(E_1,
E_2)$ satisfying the trivial constraints $\abs{E_x} \leq 1$ belong to
$\mathcal{Q}$. Indeed, let $\ket{0}$ be the ground state of $H$ and $\ket{1}$
an eigenstate with eigenvalue $1$. Clearly, $H_1 + H_2 = \Tr[H (\phi_{1} +
\phi_{2})] = 1$ can be obtained for any two orthogonal states
$\ket{\phi_{1}}$ and $\ket{\phi_{2}}$ in the space spanned by $\ket{0}$ and
$\ket{1}$. Therefore, one cannot exclude that S emits two \emph{orthogonal}
pure states $\phi_{1}$ and $\phi_{2}$. But in this case, they can encode
faithfully the two values $x = 1, 2$ and any correlations $(E_1, E_2)$ are
possible, for instance by setting $M$ to $M = E_1 \phi_1 + E_2 \phi_2$.

However, if $H_{1} + H_{2} < 1$, then intuitively $H_1$ and $H_2$ are both
small and both close to the non-degenerate ground state $\ket{0}$ of
$H$. Thus they should have a non-zero overlap which will restrict the set of
possible correlations $(E_1, E_2)$. This intuition is made precise by the
following result.

For $H_1, H_2 \geq 0$ and $H_1 + H_2 \leq 1$, the set $\setQ_{H_1, H_2}$
consists of the values $(E_{1}, E_{2})$ satisfying $\abs{E_{x}} \leq 1$ and
\begin{equation}
  \label{eq:s_charact}
  g(E_{1}, E_{2}) \geq h(H_{1}, H_{2}) \,,
\end{equation}
where
\begin{IEEEeqnarray}{rCl}
  g(E_1, E_2) &=&
  \frac{1}{2} \Bigro{\sqrt{1 + E_1} \sqrt{1 + E_2}
    + \sqrt{1 - E_1} \sqrt{1 - E_2}} \label{eq:g} \,, \\
  h(H_1, H_2) &=&
  \sqrt{1 - H_1} \sqrt{1 - H_2}-\sqrt{H_1} \sqrt{H_2} \,.
  \label{eq:h}
\end{IEEEeqnarray}

The trivial constraints $\abs{E_x} \leq 1$ follow immediately from the
definition of these quantities, so we only need to establish
\eqref{eq:s_charact}. We do this in two steps. First of all, given two pure
states $\phi_1$, $\phi_2$, the set of correlations
$(E_1, E_2) = \bigro{\Tr[M \phi_{1}], \Tr[M \phi_{2}]}$ that can be obtained
using an arbitrary measurement $M$ obviously only depends on the scalar
product $\babs{\braket{\phi_{1}}{\phi_{2}}}$. We show in
Subsection~\ref{sec:g} that this set is completely characterized by the
constraint
\begin{equation}
  \label{eq:gex}
  g(E_{1}, E_{2})\geq \babs{\braket{\phi_{1}}{\phi_{2}}} \,.
\end{equation}
We then show in Subsection~\ref{sec:h} that the parameters $H_1$, $H_2$ imply
a tight lower bound on the scalar product,
\begin{equation}
  \label{eq:gamma_h0h1_ineq}
  \babs{\braket{\phi_{1}}{\phi_{2}}} \geq h(H_{1}, H_{2}) \,.
\end{equation}
Combining these two bounds we obtain the relation $g(E_{1}, E_{2}) \geq
h(H_{1}, H_{2})$ in \eqref{eq:s_charact}.

Note that we characterize the set $\setQ$ only for values of $H_1$, $H_2$
such that $H_1 + H_2 \leq 1$. Indeed, in the next sections we are going to
use pure-state correlations in $\setQ$ as building blocks for more general
sets of correlations but under the assumption that the possible expectation
values $H_x$ of the observable $H$ are \emph{upper bounded} by some given
thresholds $\omega_x$, as in \eqref{eq:bounds} and \eqref{eq:bounds2}.  But
since any correlations $(E_1, E_2)$ are already possible in the case
$H_1 + H_2 = 1$, as we pointed out above, there is obviously no advantage in
considering larger values $H_1 + H_2 > 1$ to comply with the assumed
thresholds. From now on, we thus always consider that $H_1 + H_2 \leq 1$ (and
similarly that $\omega_1 + \omega_2 \leq 1$ in bounds of the type
\eqref{eq:bounds} and \eqref{eq:bounds2}).

\subsection{Characterization of the possible correlations $(E_1, E_2)$ for
  pure states as a function of their scalar product
  $\babs{\braket{\phi_{1}}{\phi_{2}}}$}
\label{sec:g}

Since $\abs{E_x} \leq 1$, the region of possible values $(E_1, E_2)$ is
obviously contained in the square $[-1,1]\times[-1,1]$. We now show how a
promise on the scalar product
$\babs{\braket{\phi_{1}}{\phi_{2}}} = \gamma$ further constrains the
possible values of $(E_1, E_2)$.

The parameter $\gamma$ satisfies $0\leq \gamma\leq 1$. The two extreme
cases are readily solved. When $\gamma =
\babs{\braket{\phi_{1}}{\phi_{2}}} = 1$, the two states are
indistinguishable and the measurement statistics must necessarily yield $E_1
= E_2$. When $\gamma = 0$, the two states are orthogonal and thus can
perfectly encode the value of the input $x$. In particular, we can attain any
value $(E_1, E_2) \in [-1, 1] \times [-1, 1]$ by setting $M$ to $M = E_{1}
\phi_{1} + E_{2} \phi_{2}$. These two situations can be summarized by the
relation $g(E_1, E_2) \geq \gamma$, which implies $E_1 = E_2$ when
$\gamma = 1$ and which does not put any restriction on $(E_1, E_2)$ when
$\gamma = 0$ since $g(E_1, E_2) \geq 0$ is always satisfied for $\abs{E_x}
\leq 1$.

Let us now assume $0 < \gamma < 1$. Obviously, we can restrict our analysis
to the two-dimensional subspace spanned by $\phi_1$ and $\phi_2$. In that
subspace, we can rewrite the states as qubit operators
$\phi_{x} = (\id + \vect{n}_{x} \cdot \sv) / 2$, where
$\sv = (\sigma_1, \sigma_2, \sigma_3)$ are the Pauli matrices and
$\norm{\vect{n}_x} = 1$. Then, a general measurement $M$ in that subspace can
be written as a convex combination
$(p_{0} - p_{1}) \id + p_{2} \, \vect{m} \cdot \sv$, where $p_i \geq 0$,
$\sum_i p_i = 1$, $\norm{\vect{m}} = 1$, and $\vect{m}$ can be taken in the
span of $\{\vect{n}_1, \vect{n}_2\}$ without loss of generality. The
resulting correlations correspond to the mixture
\begin{equation}
(E_1, E_2) = p_0 (1, 1) + p_1(-1, -1) + p_2(\vect{n_1} \cdot \vect{m},
\vect{n_2} \cdot \vect{m}).
\end{equation}
In other words, the region of allowed
$(E_1, E_2)$ is the convex hull of the points
\begin{equation}
\bigbr{(1,1), (-1,-1), (\vect{n_1} \cdot \vect{m}, \vect{n_2} \cdot
  \vect{m})}
\end{equation}

\begin{figure}[tbp]
  \centering
  \includegraphics{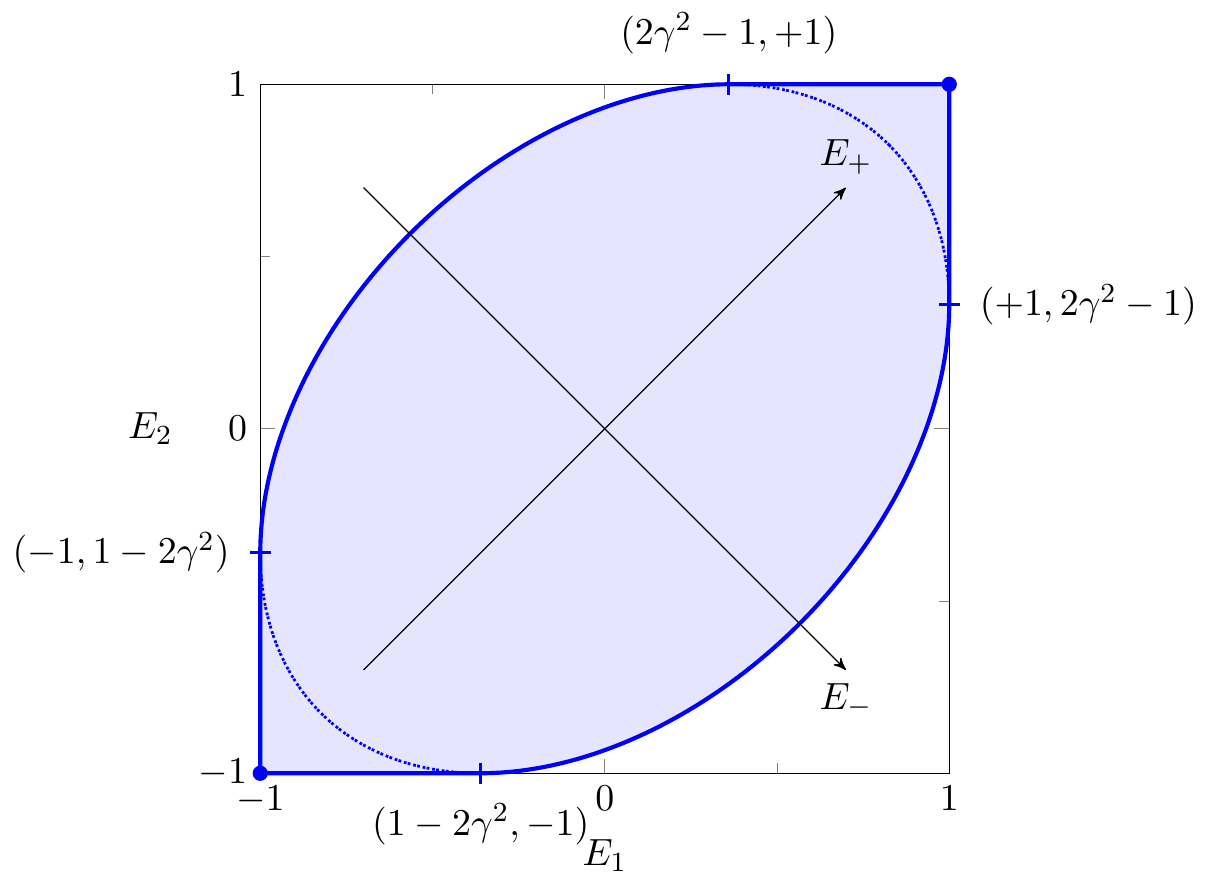}
  \caption{Representation of the ellipse \eqref{eq:BI} in the space
    $(E_1, E_2)$, depicted here for $\gamma = 0.82$. The region of
    physically possible $(E_1, E_2)$ corresponds to the convex hull of this
    ellipse with the two corner points $(1, 1)$ and $(-1, -1)$. The ellipse
    intersects the borders of the region $[-1, 1] \times [-1, 1]$ at the four
    depicted points. This defines two arcs of ellipses, corresponding to the
    portions of the ellipse represented in bold,
	and given by the solutions to $g(E_1, E_2) = \gamma$. The set
    of physically possible $(E_1, E_2)$ then corresponds to the subset of
    $[-1, 1] \times [-1, 1]$ lying between these two arcs of ellipses. This
    corresponds to the region $g(E_1, E_2) \geq \gamma$.}
\label{fig:ellipse}
\end{figure}

Let us characterize further the points
$(E_1, E_2) = (\vect{n_1} \cdot \vect{m}, \vect{n_2} \cdot \vect{m})$. Since
$\vect{m}$ lies in the span of $\vect{n}_1$, $\vect{n}_2$ and since
$(\vect{n}_1 + \vect{n}_2) \cdot (\vect{n}_1 - \vect{n}_2) = 0$, we can write
without loss of generality
\begin{equation}
  \vect{m} = \cos(\theta) \frac{\vect{n}_1 + \vect{n}_2}{
    \norm{\vect{n}_1 + \vect{n}_2}}
  + \sin(\theta) \frac{\vect{n}_1 - \vect{n}_2}{
    \norm{\vect{n}_1 - \vect{n}_2}} \,.
\end{equation}
Using this formulation together with $\norm{\vect{n}_1 + \vect{n}_2} =
2\gamma$ and $\norm{\vect{n}_1 - \vect{n}_2} = 2 \sqrt{1 - \gamma^2}$, we
find
\begin{IEEEeqnarray}{rCcCl}
  \frac{E_1 + E_2}{2 \gamma} &=& \frac{\vect{n_1} +
    \vect{n_2}}{\norm{\vect{n}_1 + \vect{n}_2}}
  \cdot \vect{m}
  &=& \cos(\theta) \,, \\
  \frac{E_1 - E_2}{2 \sqrt{1 - \gamma^2}}
  &=& \frac{\vect{n_1} - \vect{n_2}}{\norm{\vect{n}_1 - \vect{n}_2}}
  \cdot \vect{m}
  &=& \sin(\theta) \,.
\end{IEEEeqnarray}
We thus find that the set of points
$(E_1, E_2) = (\vect{n_1} \cdot \vect{m}, \vect{n_2} \cdot \vect{m})$ for a
given $\gamma$ corresponds to the ellipse
\begin{equation}
  \label{eq:BI}
  \biggro{\frac{E_+}{2 \gamma}}^2
  + \biggro{\frac{E_-}{2 \sqrt{1 - \gamma^2}}}^2 = 1 \,,
\end{equation}
where we have defined $E_\pm = E_1 \pm E_2$. The region of allowed
$(E_1, E_2)$ for an arbitrary measurement $M$ is therefore the convex hull of
$(1, 1)$, $(-1, -1)$, and any points on this ellipse, as represented in
Fig.~\ref{fig:ellipse}.

We can represent this region in a compact way as the condition
$g(E_1, E_2) \geq \gamma$. Indeed, first note that the ellipse \eqref{eq:BI}
intersects the borders of $[-1, 1] \times [-1, 1]$ at the two points
$(E_1, E_2) \in \bigbr{(2 \gamma^2 - 1, 1), (-1, 1 - 2\gamma^2)}$ in the
region above the $E_+$-axis and at the two points
$\bigbr{(1 - 2 \gamma^2, -1), (1, 2 \gamma^2 - 1)}$ in the region below the
$E_+$-axis, as represented in Fig.~\ref{fig:ellipse}. These two pairs of
points define two arcs of ellipses, as illustrated in
Fig.~\ref{fig:ellipse}. After some basic algebra (corresponding to writing
\eqref{eq:BI} explicitly in terms of $E_1$, $E_2$), one finds that these two
arcs of ellipses correspond to the values of $(E_1, E_2)$ which solve
$g(E_1, E_2) = \gamma$. It is not difficult to observe that any point in the
convex hull of $(1, 1)$, $(-1, -1)$, and the ellipse \eqref{eq:BI} also
belongs to the arcs of a second ellipse satisfying
$g(E_1, E_2) = \tilde{\gamma} \geq \gamma$. Indeed, by increasing
$\tilde{\gamma}$, the four intersection points defined above move towards the
corners $(-1, -1)$ and $(1, 1)$, which they reach when $\tilde{\gamma} = 1$
and the ellipse becomes the line segment between the points $(1, 1)$ and
$(-1, -1)$. Therefore, as $\tilde{\gamma}$ increases, the two corresponding
arcs of ellipses stretch and move over the entire convex region for
$(E_1, E_2)$ defined above. We deduce that this convex region is given by the
values of $(E_1, E_2)$ satisfying $g(E_1, E_2) \geq \gamma$.

\begin{figure}[tbp]
  \centering
  \includegraphics{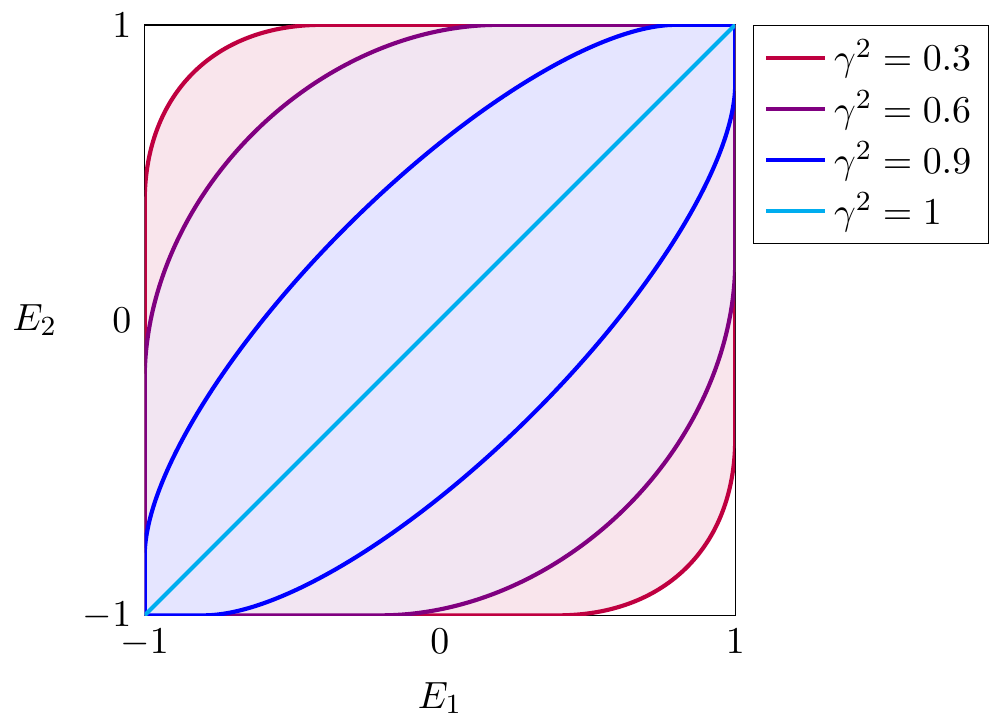}
  \caption{Region of possible values $(E_{1}, E_{2})$ satisfying
    $g(E_{1}, E_{2}) \geq \gamma$ for $\gamma = 0.55,0.77,0.95,1$.}
  \label{fig:g}
\end{figure}
In summary, we have established that the set of correlations $(E_1, E_2)$
that can be obtained by measuring two pure states with given scalar product
$\gamma = \babs{\braket{\phi_{1}}{\phi_{2}}}$ is given by
$g(E_1, E_2) \geq \gamma$, i.e., \eqref{eq:gex}. The corresponding regions
for different values of $\babs{\braket{\phi_{1}}{\phi_{2}}} = \gamma$ are
represented in Fig.~\ref{fig:g}. Note that the regions strictly grow when the
scalar product $\gamma$ decreases, starting from the line segment $E_1 = E_2$
when $\gamma = 1$ (corresponding to two indistinguishable states), to the
full square when $\gamma = 0$ (corresponding to two fully orthogonal
states).

\subsection{Lower bound on the scalar product
  $\babs{\braket{\phi_{1}}{\phi_{2}}}$ as a function of expectation values
  $(H_1, H_2)$}
\label{sec:h}

We have just seen that the region of possible correlations $(E_1, E_2)$
attainable with pure states only depends on their scalar product through
\eqref{eq:gex}. We now show that the parameters $(H_1, H_2)$ constrain the
possible values of this scalar product through the lower bound
$\babs{\braket{\phi_{1}}{\phi_{2}}} \geq h(H_{1}, H_{2})$ where
$h(H_1, H_2)$ is defined in \eqref{eq:h}. Intuitively if $H_1$ and $H_2$ are
small, both states $\phi_1$ and $\phi_2$ are close to the non-degenerate
ground state $\ket{0}$ of $H$ and thus they should have a non-zero
overlap. This is what \eqref{eq:h} makes precise.

Let us first show that for any values of $(H_1, H_2)$ it is indeed possible
to find two states $\phi_1$, $\phi_2$ such that their scalar product
satisfies $\babs{\braket{\phi_{1}}{\phi_{2}}}=h(H_{1}, H_{2})$. This
implies that any point $(E_{1}, E_{2})$ satisfying \eqref{eq:s_charact} can
indeed be realized in our prepare-and-measure scenario.

For this, simply note that any $H_{1}$ and $H_{2}$ satisfying $H_{1}, H_{2}
\geq 0$ and $H_{1} + H_{2} \leq 1$ can be expressed as $H_x =
\sin(\theta_x)^{2}$ for some suitable $\theta_x \in [0, \pi/2]$. Computing
$h(H_{1}, H_{2})$, we find
$h(H_{1}, H_{2}) = \cos(\theta_1 + \theta_2)$. Define the two states
\begin{IEEEeqnarray}{rCl}
  \ket{\phi_1} &=& \cos(\theta_1) \ket{0} + \sin(\theta_1) \ket{1} \,, \\
  \ket{\phi_2} &=& \cos(\theta_2) \ket{0} - \sin(\theta_2) \ket{1} \,.
\end{IEEEeqnarray}
Then, their scalar product satisfies $\babs{\braket{\phi_{1}}{\phi_{2}}}
= \cos(\theta_1 + \theta_2) = h(H_1, H_2)$, as required.

Let us now show that the scalar product between $\phi_1$, $\phi_2$ cannot be
smaller than the value given by \eqref{eq:h}. This implies
that no correlations outside the region defined by \eqref{eq:s_charact} can
be obtained by measuring pure states in our scenario.

For this, let $\beta_x = 1 - \bra{0} \phi_x \ket{0}$ be the weight of the
state $\phi_x$ in the subspace orthogonal to the groundstate $\ket{0}$ of
$H$. We have that $\beta_x\leq H_x$ since
$H_x = \Tr[H \phi_{x}] = \sum_{i>0} \lambda_i \Tr[P_i \phi_x] \geq \sum_{i>0}
\Tr[P_i \phi_x] = 1 - \Tr[P_0\phi_x] = \beta_x$, where $\lambda_i$ and $P_i$
denote the eigenvalues of $H$ and the corresponding projectors, with
$\lambda_0 = 0$ and $\lambda_i \geq 1$ for $i > 0$. Writing
\begin{equation}
  \ket{\phi_x} = \sqrt{1 - \beta_x} \ket{0} + \sqrt{\beta_x} \ket{\xi_x} \,,
\end{equation}
where $\ket{\xi_x}$ is in the subspace orthogonal to $\ket{0}$, we have that
\begin{IEEEeqnarray}{rCl}
  \abs{\braket{\phi_1}{\phi_2}}
  &=& \babs{\sqrt{1-\beta_1} \sqrt{1-\beta_2}
    + \sqrt{\beta_1} \sqrt{\beta_2} \braket{\xi_1}{\xi_2}}
  \IEEEnonumber \\
  &\geq& \babs{\sqrt{1 - \beta_1} \sqrt{1 - \beta_2}
    - \sqrt{\beta_1} \sqrt{\beta_2}} \IEEEnonumber \\
  &=& \sqrt{1 - \beta_1} \sqrt{1 - \beta_2}
  - \sqrt{\beta_1} \sqrt{\beta_2} \IEEEnonumber \\
  &=& h(\beta_1, \beta_2) \,,
\end{IEEEeqnarray}
where in the second line we used that the right hand-side is minimized when
$\braket{\xi_1}{\xi_2} = -1$ and in the last line we used that
$h(\beta_1, \beta_2)$ is positive when $\beta_1 + \beta_2 \leq 1$, which is
the case since $H_1 + H_2 \leq 1$ and $\beta_x \leq H_x$. Now, given that the
function $h(x, y)$ is non-decreasing in $x$ and $y$ in the region
$x + y \leq 1$, we have that $h(\beta_1, \beta_2) \geq h(H_1, H_2)$, which
gives \eqref{eq:h}.

\section{General quantum correlations}
\label{sec:genq}

\subsection{Shared randomness}

Let us now turn to the general case where S and M can exploit shared random
parameters $\lambda$. This includes in particular the case where the source S
can emit \emph{mixed} states since any mixed state $\rho_x$ can be viewed as
a convex decomposition $\rho_x = \sum_\lambda p_\lambda \phi_{x}^\lambda$ of
pure states. Let
\begin{equation}
  \label{eq:Qm}
  \setQ'_{H_1, H_2} = \Bigbr{
    \vect{E} = \sum_\lambda p_\lambda \vect{E}_\lambda
    \bigm| \vect{E}_\lambda \in \setQ_{\vect{H}_\lambda} , \,
    \sum_\lambda p_\lambda \vect{H}_\lambda = \vect{H}} \,,
\end{equation}
be the set of possible quantum correlations $(E_1, E_2)$ compatible with
given expectation values $(H_1, H_2)$ in the presence of shared randomness,
where we write $\vect{E} = (E_1, E_2)$ for the correlations averaged over the
shared randomness $\lambda$,
$\vect{E}_\lambda = (E_{1|\lambda}, E_{2|\lambda})$ for the correlations
corresponding to a specific value of the shared randomness, and similarly for
$\vect{H}$ and $\vect{H}_\lambda$. The set $\setQ'$ (as in the previous
section we drop the subindices $H_1$, $H_2$ to simplify the notation) thus
corresponds to convex sums
$\vect{E} = \sum_\lambda p_\lambda \vect{E}_\lambda$ of pure-state quantum
realizations, each of which is characterized by expectation values
$\vect{H}_\lambda$. The values $\vect{E}_\lambda$ must thus belong to
$\setQ_{\vect{H}_\lambda}$ and we require in addition that we recover on
average the given expectations $\vect{H}$:
$\sum_\lambda p_\lambda \vect{H}_\lambda=\vect{H}$.

We now show that
\begin{equation}
  \setQ'_{H_1, H_2} = \setQ_{H_1, H_2} \,,
\end{equation}
i.e., the set of pure-state correlations is closed under
shared-randomness. Interestingly, the same property is also shared by
pure-state quantum correlations in the context of Bell non-locality, but is
not true in the context of semi-DI scenarios based on dimension bounds.
	
Since obviously $\setQ' \supseteq \setQ$, we need only prove that
$\setQ' \subseteq \setQ$, i.e., that given any mixture
$\vect{E} = \sum_\lambda p_\lambda \vect{E}_\lambda$ of pure-state
correlations with average expectations
$\vect{H} = \sum_\lambda p_\lambda \vect{H}_\lambda$, then the exact same
correlations $\vect{E}$ can be obtained through a single pure-state quantum
realization with expectation values $\vect{H}$. This simply amounts to
showing that $g(\vect{E})\geq h(\vect{H})$ since this condition fully
characterizes the set of pure-state quantum correlations, or equivalently
that $g(\vect{E})^{2} - h(\vect{H})^{2} \geq 0$ since the functions $g$ and
$h$ are positive in the domain $H_1+H_2\leq 1$. We establish this by showing
that $g^2$ and $-h^2$ are concave. Indeed, if this is the case we have
\begin{IEEEeqnarray}{rCl}
  g(\vect{E})^{2} - h(\vect{H})^{2}
  &=& g \bigro{\textstyle \sum_{\lambda} p_{\lambda} \vect{E}_{\lambda}}^{2}
  - h \bigro{\sum_{\lambda} p_{\lambda} \vect{H}_{\lambda}}^{2}
  \IEEEnonumber \\
  &\geq& \sum_{\lambda} p_{\lambda} g(\vect{E}_{\lambda})^{2}
  - \sum_{\lambda} p_{\lambda} h(\vect{H}_{\lambda})^{2}
  \IEEEnonumber \\
  &=& \sum_{\lambda} p_{\lambda} \bigro{
    g(\vect{E}_{\lambda})^{2} - h(\vect{H}_{\lambda})^{2}} \IEEEnonumber \\
 &\geq& 0
\end{IEEEeqnarray}
where we used the concavity of $g^2$ and $-h^2$ in the first inequality and
the condition $g(\vect{E}_{\lambda})^{2} - h(\vect{H}_{\lambda})^{2} \geq 0$
for each $\lambda$ in the second inequality, since by assumption
$\vect{E}_\lambda \in \mathcal{Q}_{\vect{H}_\lambda}$.

Note that both $g^2$ and $-h^2$ can be written in term of the function
\begin{equation}
  f(x, y) = \Bigro{\sqrt{x y} + \sqrt{(1 - x) (1 - y)}}^{2}
\end{equation}
as $g(E_1, E_2)^{2} = f\bigro{\frac{1 + E_1}{2}, \frac{1 + E_2}{2}}$ and
$-h(H_1, H_2)^{2} = f(H_1, 1 - H_2) - 1$. Showing the concavity of $g^2$ and
$-h^2$ thus reduces to showing that $f$ is concave for $0 \leq x, y \leq 1$,
i.e., that its Hessian matrix $\hess(f)$ is negative semidefinite. A
straightforward computation shows that $\Tr \bigsq{\hess(f)} \leq 0$ and
$\det \bigsq{\hess(f)} \geq 0$ for any $0 \leq x, y \leq 1$. Since $\hess(f)$
is a $2 \times 2$ symmetric matrix, this implies, as desired, that both of
its eigenvalues are nonpositive.

\subsection{Upper bounds on the expectation values of $H$}

The sets $\setQ$ and $\setQ'$ defined above assume that the source S emits
states with given expectation values $\vect{H} = (H_1, H_2)$ for the
observable $H$. However, as we pointed out in Section~\ref{sec:frame}, rather
than assuming that $H$ takes some exact values, it is more natural to assume
upper bounds on the possible values of $H$, and we introduced two possible
ways to bound such values, either through the max-average assumption
\eqref{eq:bounds} or the max-peak assumption \eqref{eq:bounds2}. This leads
us to define the set of quantum correlations under the max-average assumption
as
\begin{equation}
  \label{eq:Qa}
  \setQa_{\h_1,\h_2} = \Bigbr{
    \sum_\lambda p_{\lambda} \vect{E}_\lambda
    \mid \vect{E}_\lambda \in \setQ_{\vect{H}_\lambda}, \,
    \sum_\lambda p_\lambda \vect{H}_\lambda\leq \vect{\h}}
\end{equation}
and the set of quantum correlations under the max-peak assumption as
\begin{equation}
  \label{eq:Qp}
  \setQp_{\h_1,\h_2} = \Bigbr{
    \sum_\lambda p_{\lambda} \vect{E}_\lambda
    \mid \vect{E}_\lambda \in \setQ_{\vect{H}_\lambda}, \,
    \max_\lambda \vect{H}_{\lambda} \leq \vect{\h}} \,,
\end{equation}
where we have written $\vect{\h} = (\h_1, \h_2)$. Note that following the
remark made at the end of Section~\ref{sec:Q}, we always assume that
$\h_1 + \h_2 \leq 1$, since any possible correlations $(E_1, E_2)$ are already
possible in the case $\h_1 + \h_2 = 1$.

From our previous results, the characterization of $\setQa_{\h_1, \h_2}$ and
$\setQp_{\h_1, \h_2}$ is immediate. First, it is easy to check that the sets
$\setQ_{H_1, H_2}$ of allowed values of $(E_1, E_2)$ are strictly increasing
with $H_1$, $H_2$. That is, $\setQ_{H_1, H_2} \subseteq \setQ_{H'_1, H'_2}$
if $H_1 \leq H'_1$ and $H_2 \leq H'_2$. This follows from the fact that
$h(H_1, H_2)$ is a non-increasing function of $H_1$ and $H_2$ in the range
$H_1 + H_2 \leq 1$ and that the sets of values $(E_1, E_2)$ defined by
$g(E_1, E_2) \geq \gamma$ increase with decreasing $\gamma$, as
illustrated in Fig.~\ref{fig:g}. It thus follows that in \eqref{eq:Qa} one
can always assume that
$\sum_\lambda p_\lambda \vect{H}_{\lambda} = \vect{\h}$ and in \eqref{eq:Qp}
that $\vect{H}_{\lambda} = \vect{\h}$ for all values of $\lambda$. Using the
results of the previous subsection on the behavior of $\setQ$ under shared
randomness, we deduce that
\begin{equation}
  \setQa_{\h_1, \h_2} = \setQ'_{\h_1, \h_2}
  = \setQ_{\h_1, \h_2} = \setQp_{\h_1, \h_2} \,.
\end{equation}
We thus conclude that any correlations that can be generated by an arbitrary
quantum realization constrained only by upper bounds $\h_1$ and $\h_2$ on
$H_{1}$ and $H_{2}$, whether through the max-average or the max-peak
assumption, always admit an equivalent pure-state quantum representation
saturating these upper bounds. In this sense the max-average and max-peak
assumptions are equivalent. As we will see in Sections~\ref{sec:c} and
\ref{sec:d}, however, different quantum realizations of the same correlations
may exhibit different underlying quantum properties, and the max-average and
max-peak assumptions are different from this perspective.

\section{Classical correlations}
\label{sec:c}

A basic property of fully- or semi-DI setups based on non-locality or
dimension bounds is the existence of quantum correlations that have no
classical analogue. This is clearly a prerequisite for any application of
such correlations, e.g., for randomness certification. This property is also
present in our semi-DI scenario as we now show.

\subsection{Definition}

We need first to define some notion of ``classicality'' in our context and a
corresponding set of correlations. The no-communication assumption in
standard Bell tests and the dimension bound in usual semi-DI protocols have a
well-defined meaning in a classical context, without any reference to quantum
theory. This is no longer the case for the assumptions that we consider here,
as they are expressed as constraints on the mean value of a quantum
observable $H$ and hence explicitly assume some underlying quantum model. It
is nevertheless still possible to identify sets of correlations that are
``classical'' in the sense that they do not exhibit genuinely quantum
features and thus are useless for semi-DI applications. The most
straightforward way to do so is to proceed by analogy with standard Bell
tests or usual semi-DI protocols, where classical correlations correspond
mathematically to those that can be expressed, with the help of shared
randomness, as convex combinations of \emph{deterministic} correlations.

We thus define the set of classical correlations under the max-average
assumption \eqref{eq:bounds} as
\begin{equation}
  \label{eq:c_max_avg}
  \setCa_{\h_1, \h_2} = \biggbr{
    \sum_{\lambda} p_\lambda \vect{E}_{\lambda}
    \Bigm| \vect{E}_{\lambda} \in \setQ_{\vect{H}_{\lambda}} , \,
    \vect{E}_{\lambda} \in \{\pm 1\} \times \{\pm 1\} , \,
    \sum_\lambda p_\lambda \vect{H}_{\lambda} \leq \vect{\h}}
\end{equation}
and under the max-peak assumption as
\begin{equation}
  \label{eq:c_max_peak}
  \setCp_{\h_1, \h_2} = \biggbr{
    \sum_{\lambda} p_\lambda \vect{E}_{\lambda}
    \Bigm| \vect{E}_{\lambda} \in \setQ_{\vect{H}_{\lambda}} , \,
    \vect{E}_{\lambda} \in \{\pm 1\} \times \{\pm 1\} , \,
    \max_\lambda \vect{H}_{\lambda} \leq \vect{\h}} \,.
\end{equation}
The constraint $\vect{E}_{\lambda} \in \{\pm 1\} \times \{\pm 1\}$ in these
definitions implies that for any given value of the shared randomness
$\lambda$, an output $\pm 1$ for the measurement performed at M is completely
pre-determined for each of the two states emitted by the source S\@. Thus no
genuinely quantum behavior is exhibited by the two devices. Conversely, if
some correlations $\vect{E}$ lie outside the set $\setCa_{\h_1, \h_2}$, then
necessarily the output of M cannot be predetermined for at least one of the
states sent by S, a typically quantum feature.

We show that $\setCa_{\h_1, \h_2}$ is a polytope, which apart from the
trivial facets $\abs{E_x} \leq 1$ is defined by
\begin{equation}\label{eq:class_constraint}
  \abs{E_-} = \abs{E_1 - E_2} \leq 2 (\h_{1} + \h_{2})\,.
\end{equation}
Similarly, $\setCp_{\h_1, \h_2}$ is a polytope characterized by the stronger
inequality
\begin{equation}\label{eq:cpeak}
  \abs{E_{-}} = \abs{E_1 - E_2} \leq 2 \, \Theta(\h_1 + \h_2) \,,
\end{equation}
where $\Theta(z) = 0$ if $z < 1$ and $\Theta(z) = 1$ if $z = 1$.

Let us first establish the characterization \eqref{eq:class_constraint} of
$\setCa$. Remark that for any
$(\vect{E}_{\lambda}, \vect{H}_{\lambda}) \in \setQ$ for which
$\vect{E}_{\lambda}\in\{\pm 1\}\times\{\pm 1\}$ there are two
possibilities. Either $E_{1|\lambda} = E_{2|\lambda}$, in which case
$\abs{E_{1|\lambda} - E_{2|\lambda}} = 0$. Or
$E_{1|\lambda} = - E_{2|\lambda}$, in which case the states emitted for
$x = 1$ and $x = 2$ must be orthogonal pure states and thus
$H_{1|\lambda} + H_{2|\lambda} \geq 1 = \abs{E_{1|\lambda} -
  E_{2|\lambda}}/2$. By taking convex combination of these possibilities, we
find
\begin{equation}
  \abs{E_1 - E_2}
  \leq \sum_\lambda p_\lambda \abs{E_{1|\lambda} - E_{2|\lambda}}
  \leq \sum_\lambda p_\lambda 2(H_{1|\lambda} + H_{2|\lambda})
  \leq 2(\h_{1} + \h_{2}) \,.
\end{equation}

Conversely, any correlations $\vect{E}$ satisfying the constraint
\eqref{eq:class_constraint} belong to $\setCa$. To prove this, note that the
polytope defined by \eqref{eq:class_constraint} has the following six extreme
points
\begin{equation}
  (E_1, E_2) = \Bigbr{\bigro{\pm 1, \pm 1} , \,
    \bigro{\pm 1, \pm[1 - 2(\h_1 + \h_2)]} , \,
    \bigro{\pm[1 - 2(\h_1 + \h_2)], \pm 1}} \,.
\end{equation}
Proving that any $\vect{E}$ in this polytope belongs to $\setCa$ amounts to
showing, by convexity, that any of these six extreme points belongs to
$\setCa$. This is evident for the two points $(\pm 1, \pm 1)$. The point
$\bigro{1,1-2(\h_1+\h_2)}$ can be decomposed as
$p_1 \vect{E}_1+p_2\vect{E}_2+p_3\vect{E}_3$, where
\begin{IEEEeqnarray}{rCl+rCl}
  p_1 &=& 1 - \h_1 - \h_2 \,, & \vect{E}_1 &=& (1,1) \in \setQ_{0,0} \,, \\
  p_2 &=& \h_1 \,, & \vect{E}_2 &=& (1,-1) \in \setQ_{1,0} \,, \\
  p_3 &=& \h_2 \,, & \vect{E}_3 &=& (1,-1) \in \setQ_{0,1} \,,
\end{IEEEeqnarray}
and thus also belongs to $\setCa$. Similar decompositions are readily
obtained for the three other points.

The characterization \eqref{eq:cpeak} of $\setCp$ follows from two simple
observations. If $\h_1 + \h_2 < 1$, the only points
$\vect{E}_\lambda \in \setQ$ such that
$\vect{E}_\lambda \in \{\pm 1\} \times \{\pm 1\}$ are $(1,1)$ and $(-1,
-1)$. Their convex combination defines the line segment $E_1 - E_2 = 0$. If
$\h_1 + \h_2=1$, the four corner points $\{\pm 1\} \times \{\pm 1\}$ are
available, and as usual we have no constraints on $\vect{E}$, so that
$\abs{E_1 - E_2}$ can reach the maximal value 2.

\subsection{Bell inequality analogues}

\begin{figure}[tbp]
  \centering
  \includegraphics{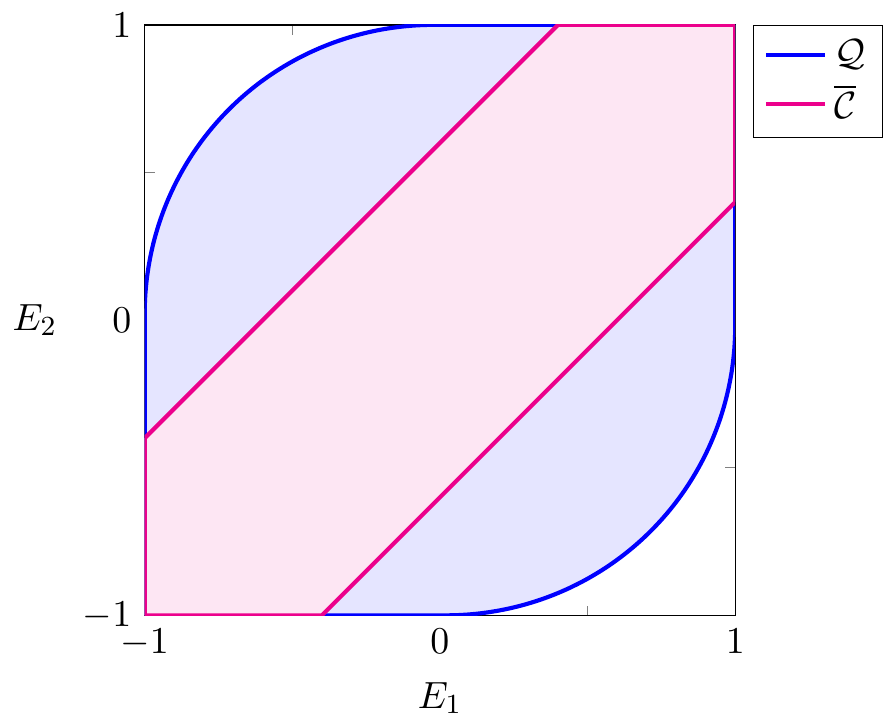}
  \caption{Comparison of the quantum set $\setQ$ and the classical set
    $\setCa$ for $\h_{1} = \h_{2} = 0.15$, showing that it is possible to
    violate the classical bound \eqref{eq:class_constraint} with quantum
    systems. The $E_-$-axis ($E_- = E_1 - E_2)$ measures how quantum the
    behavior of the boxes is, in a way similar to the CHSH witness in the
    study of non-locality. For the particular case $\h_{1} = \h_{2} = 0.15$,
    classical systems are limited to $\abs{E_-}\leq 0.6$ but the maximal
    value for quantum systems is $\abs{E_-} \simeq 1.43$.\label{fig:c}}
\end{figure}

The inequalities \eqref{eq:class_constraint} and \eqref{eq:cpeak} play the
same role as the inequality $\abs{\text{CHSH}} \leq 2$ in the context of Bell
non-locality, in that they separate the quantum region from the region of
convex combinations of deterministic correlations, see illustration of the
sets $\setCa$ and $\setQ$ in Fig.~\ref{fig:c}. The analogue of the Tsirelson
inequality $\abs{\text{CHSH}} \leq 2 \sqrt{2}$ is the inequality
$\abs{E_-} = \abs{E_1 - E_2} \leq 2 \bigro{\sqrt{H_1} \sqrt{1 - H_2} +
  \sqrt{1 - H_1} \sqrt{H_2}}$. This quantum bound is readily obtained from
the results of the previous sections (the points maximizing $\abs{E_1 - E_2}$
in $\setQ$ correspond to the points on the $E_-$-axis of the ellipse
\eqref{eq:BI}).

The axis $E_-$ thus corresponds to the axis along which ``quantumness''
increases. Note that we can interpret $\abs{E_-}$ as a measure of how well it
is possible to guess which input $x=1$ or $x=2$ was used given the
measurement outcome $b$. Indeed, since
\begin{equation}
  \tfrac{1}{2} \abs{E_-} = \tfrac{1}{2} \abs{E_{1} - E_{2}}
  = \tfrac{1}{2} \sum_{b = \pm 1} \babs{P(b|x=1) - P(b|x=2)} \,,
\end{equation}
the quantity $\tfrac{1}{2} \abs{E_-}$ is equal to the statistical distance
\begin{equation}
  d\bigro{P_{B | x=1}, P_{B | x=2}}
  \equiv \tfrac{1}{2} \sum_{b = \pm 1} \babs{P(b|x=1) - P(b|x=2)}
\end{equation}
between the output probability distributions $P_{B|x=1}$ and $P_{B|x=2}$ for
the two possible inputs. In particular, the optimal probability to guess the
input is $p_\rg = \tfrac{1}{2} (1 + \tfrac{1}{2} \abs{E_-})$. For given
$H_1$, $H_2$, this probability is maximal when $\abs{E_-}$ achieves its
maximal value
$2\bigro{\sqrt{H_1} \sqrt{1 - H_2} + \sqrt{1 - H_1} \sqrt{H_2}}$, in which
case we know for sure that the two states sent by S are as distinguishable as
possible given the constraints on $H_1$ and $H_2$, and that M implements the
measurement that best distinguishes them. When
$\sqrt{H_1} \sqrt{1 - H_2} + \sqrt{1 - H_1} \sqrt{H_2} < 1$ (which
necessarily happens when $H_1 + H_2 < 1$), it follows that it is not possible
to deterministically guess which of the two choices $x = 1$ or $x = 2$ were
made on S, which proves that the quantum channel relating S to M has sub-unit
capacity, as previously anticipated.

\subsection{Implementations discussion}

The three experimental implementations that we have presented in
Subsection~\ref{sec:impl} generate non-classical correlations as illustrated
in Fig.~\ref{fig:points}. Note that the BPSK and 2ASK schemes do not admit a
fully deterministic explanation under the max-average assumption, i.e., even
if the source S is allowed to send states with arbitrary values for $H$,
provided that the average values do not exceed the assumed thresholds. This
is not true for the OOK implementation, which admits a deterministic
explanation in this case. However, such a deterministic explanation is no
longer possible if the peak values for $H$ are constrained.

Note that for $\h_1 + \h_2< 1$, the set $\setCp$ is of measure 0. We can get
some intuition for the meaning of this set as follows. If $\h_1 + \h_2 < 1$,
then for given $\lambda$ every pair of states $\phi_{1}^\lambda$,
$\phi_{2}^\lambda$ emitted by the source must be non-orthogonal, as follows
from the results of Subsection~\ref{sec:h} and the max-peak assumption. Since
there is no non-trivial measurement that will yield with certainty definite
outcomes for two non-orthogonal states, the measurement device M, if it
behaves deterministically, must actually ignore the quantum messages sent by
M and simply output a pre-registered outcome $b = -1$ or $b = 1$,
independently of whether $x = 1$ or $x = 2$. We thus necessarily have
$E_1 = E_2$ in this case. Conversely, if we observe correlations for which
$\abs{E_1 - E_2} > 0$, we can conclude that the measurement device M did not
simply output pre-registered values, but actually performed a non-trivial
measurement on the states emitted by S, which are non-orthogonal. That is,
the observation of $\abs{E_1 - E_2} > 0$ witnesses a typical quantum feature,
which in particular necessarily results in a non-deterministic outcome for at
least one of $x = 1, 2$. A similar conclusion can be reached under the
max-average assumption, but this now requires $\abs{E_1 - E_2}$ to be above a
finite value $2(\h_1 + \h_2)$, as follows from \eqref{eq:class_constraint}.

\section{Correlations exhibiting certified randomness}
\label{sec:d}

If S and M generate correlations in the classical sets $\setCa$ or $\setCp$,
then the output $b$ is predetermined simultaneously for \emph{both} choices
of inputs $x = 1, 2$, and the apparent randomness of $b$ only arises from the
pre-established classical randomness $\lambda$. Observing a point outside the
sets $\setCa$ or $\setCp$ thus guarantees that at least one of the inputs,
$x = 1$ or $x = 2$ leads to genuinely random outcomes, but does not guarantee
that a specific one, say $x = 1$ does, or that both $x = 1$ and $x = 2$
do. For instance, as can be seen in Fig.~\ref{fig:c}, the point
$(E_1, E_2) = \bigro{1, 2h^2(\h_1, \h_2) - 1}$ does not belong to $\setCa$ but
nevertheless corresponds to a situation where M returns $b=1$ with certainty
when $x=1$ is used. A similar situation arises in the context of Bell
non-locality, in which correlations can be non-local and yet have definite
values for a subset of the measurement inputs \cite{ref:bkp2006}.

In cryptographic applications, and for instance in DI or semi-DI RNG
protocols, it is usually the case that the certified randomness comes from a
fixed subset of the inputs. For instance, in our context, a semi-DI RNG
protocol along the lines of \cite{ref:pam2010,ref:pm2013,ref:nbsp2016} would
most of the time use the input $x = 1$ to generate the random string and from
time to time both $x = 1$ and $x = 2$ to estimate the correlations $\vect{E}$
produced by the devices. Given the estimated $\vect{E}$, it is then possible
to lower bound the amount of randomness extractable from the $x = 1$
measurement data.

This leads us to consider sets based on a weaker constraint than those
introduced in the previous section, those for which the output of M is
deterministic when a specified input $x$ is chosen, while potentially random
for the remaining inputs:
\begin{IEEEeqnarray}{rCl}
  \setDa_{x, \h_1, \h_2} &=& \biggbr{
    \sum_{\lambda} p_\lambda \vect{E}_{\lambda}
    \Bigm| \vect{E}_{\lambda} \in \setQ_{\vect{H}_{\lambda}} , \,
    E_{x|\lambda} \in \{\pm 1\} , \,
    \sum_\lambda p_\lambda \vect{H}_{\lambda} \leq \vect{\h}} \,,
  \label{eq:d1_av} \\
  \setDp_{x, \h_1, \h_2} &=& \biggbr{
    \sum_{\lambda} p_\lambda \vect{E}_{\lambda}
    \Bigm| \vect{E}_{\lambda} \in \setQ_{\vect{H}_{\lambda}} , \,
    E_{x|\lambda} \in \{\pm 1\} , \,
    \max_\lambda \vect{H}_{\lambda} \leq \vect{\h}} \,.
  \label{eq:d1_pk}
\end{IEEEeqnarray}
The sets $\setDa_1$, $\setDa_2$ clearly contain $\setCa$ but can be larger,
and similarly for $\setDp_1$, $\setDp_2$ with respect to $\setCp$, see
Fig.~\ref{fig:d}. Observing a point outside $\setDa_x$ or $\setDp_x$ now
certifies that the output of M is (at least to some extent) random when the
input $x$ is chosen. Furthermore, observing a point outside the convex hull
of $\setDa_1 \cup \setDa_2$ or $\setDp_1 \cup \setDp_2$ guarantees that the
output of M is random, independently of which input $x = 1$ or $x = 2$ was
used.

The sets $\setDa_x$ and $\setDp_x$ have a direct significance in the context
of semi-DI random number generation (RNG) protocols. The existence of quantum
correlations outside, say, $\setDa_1$ or $\setDp_1$ is sufficient to
guarantee the existence of a semi-DI RNG protocol along the lines of
\cite{ref:pam2010,ref:pm2013,ref:nbsp2016}, in which the input $x=1$ is used
to generate the random string and both $x = 1$ and $x = 2$ are used on a
smaller subsets of the runs to estimate the correlations $\vect{E}$. As long
as the estimated $\vect{E}$ is outside $\setDa_1$ or $\setDp_1$ (modulo
statistical corrections), one will be able to certify that a certain amount
of randomness has been produced (how to quantify precisely this randomness
will be presented in \cite{ref:vhPREP}).

\subsection{Characterization of the set $\setDa_x$}

We show here that $\setDa_x$ consists of the values $\vect{E}$ satisfying
$\abs{E_x} \leq 1$ and
\begin{IEEEeqnarray*}{rCl}
\label{eq:dall}\IEEEyesnumber \IEEEyessubnumber*
  E_x \, h^2 \biggro{\frac{2 \h_1}{1 + E_x},
    \frac{2 \h_2}{1 + E_x}} - E_{\bar{x}}
  &\leq& 1 - h^2 \biggro{\frac{2 \h_1}{E_x + 1}, \frac{2 \h_2}{E_x + 1}} \,,
  \label{eq:d1} \\
  E_x \, h^2 \biggro{\frac{2 \h_1}{1 - E_x},
    \frac{2 \h_2}{1 - E_x}} - E_{\bar{x}}
  &\geq& -1 + h^2 \biggro{\frac{2 \h_1}{1 - E_x}, \frac{2 \h_2}{1 - E_x}} \,,
  \label{eq:d2}
\end{IEEEeqnarray*}
where $\bar{x}$ denotes the input complementary to $x$ ($\bar{x} = 2$ if
$x = 1$ and $\bar{x} = 1$ if $x = 2$). 

Let us consider the case $x=1$ to simplify the notation (the derivation is the same
for $x=2$). Let $\vect{E}=\sum_{\lambda}p_\lambda \vect{E}_{\lambda}$ be an
arbitrary point in $\setDa_1$. Define $\Lambda_{\pm}$ as the set of
$\lambda$'s for which $E_{1|\lambda}=\pm 1$ and write
\begin{IEEEeqnarray}{rCl}
  \vect{E} &=& p \vect{E}_{+} + (1-p) \vect{E}_{-} \,, \\
  \vect{\h} &\geq& p \vect{H}_{+}+(1-p)\vect{H}_{-} \,,
\end{IEEEeqnarray}
where 
\begin{IEEEeqnarray}{c+c}
  p = p_{+} = \sum_{\lambda \in \Lambda_{+}} p_\lambda \,, &
  1 - p = p_{-} = \sum_{\lambda \in \Lambda_{-}} p_\lambda
\end{IEEEeqnarray}
and
\begin{IEEEeqnarray}{c+c}
  \vect{E}_{\pm} = \frac{1}{p_{\pm}}
  \sum_{\lambda \in \Lambda_{\pm}} p_\lambda \vect{E}_\lambda \,, &
  \vect{H}_{\pm} = \frac{1}{p_{\pm}}
  \sum_{\lambda \in \Lambda_{\pm}} p_\lambda \vect{H}_\lambda \,.
\end{IEEEeqnarray}

We have thus regrouped the components $(\vect{E}_\lambda, \vect{H}_\lambda)$
in two subsets $(\vect{E}_{\pm}, \vect{H}_{\pm})$, for which
$E_{1|+} = +1$ and $E_{1|-} = -1$, respectively. Since $\setQ$ is convex, the
two points $(\vect{E}_{\pm}, \vect{H}_{\pm})$ belong to $\setQ$ and satisfy
the constraints obtained in Section~\ref{sec:Q}.

In particular, since $E_{1|+} = 1$, it follows that
$E_{2|+} \geq 2 h^2 (H_{1|+}, H_{2|+}) - 1$ and thus that
\begin{equation}
  E_{2} \geq p \bigro{2 h^2 (H_{1|+}, H_{2|+}) - 1} - (1 - p)
  = p \, 2 h^2(H_{1|+}, H_{2|+}) - 1 \,.
\end{equation}
On the other hand,
\begin{equation}
  H_{1|+} \leq \frac{\h_1 - (1 - p) H_{1|-}}{p} \leq \frac{\h_1}{p} \,,
\end{equation}
and similarly $H_{2|+} \leq \h_2 / p$. It is easily established that the
function $h(x, y)$ is non increasing in its two arguments and thus that
\begin{equation}
  E_{2} \geq p \, 2  h^2 \biggro{\frac{\h_{1}}{p}, \frac{\h_{2}}{p}} - 1 \,.
\end{equation}
We can now use that $p=(1+E_1)/2$ and substitute in the inequality above,
which gives \eqref{eq:d1}. Following the same lines to lower bound $E_2$, one
obtains \eqref{eq:d2}. Finally, it is not difficult to verify that all the
intermediate inequalities in our derivation are tight and thus that any
$\vect{E}$ in the region defined by \eqref{eq:dall} can be
attained by points in $\mathcal{D}_1$. 

\begin{figure}[p]
  \centering
  \includegraphics{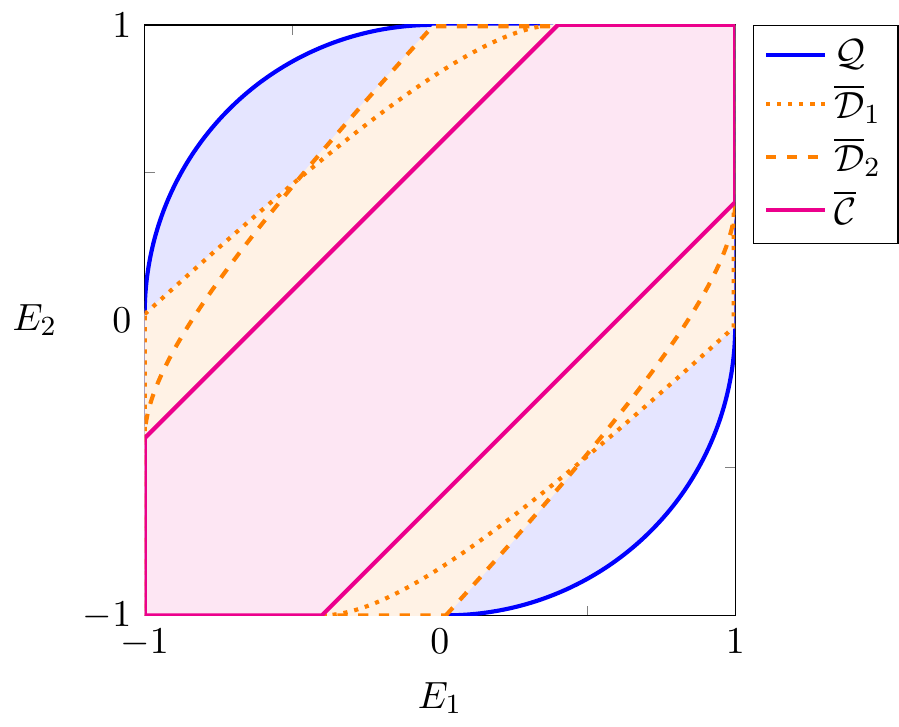}
  \caption{Representations of the sets $\setCa$, $\setDa_1$, $\setDa_2$, and
    $\setQ$ for $\h_{1} = \h_{2} = 0.15$. Since the quantum region is
    strictly larger than the individual sets $\setDa_x$ (or even than their
    convex combination), it is possible to certify the production of genuine
    randomness. \label{fig:d}}
\end{figure}

The sets $\setDa_x$ are compared in Fig.~\ref{fig:d} to $\setCa$ and
$\setQ$.

\subsection{Characterization of the set $\setDp_x$}

The set $\setDp_x$ consists of the values $\vect{E}$ satisfying
$\abs{E_x} \leq 1$ and
\begin{equation}\label{eq:dpeak}
  \babs{E_x \, h^2(\h_1, \h_2) - E_{\bar x}} \leq 1 - h^2(\h_1, \h_2) \,.
\end{equation}

In order to establish the formula \eqref{eq:dpeak}, note that, if
$E_{x|\lambda} = 1$ and $H_{x|\lambda} \leq \h_x$, then necessarily
$1 \geq E_{\bar x, \lambda} \geq 2h^2(\h_1, \h_2) - 1$, while if
$E_{x, \lambda} = -1$, it holds that
$-1 \leq E_{\bar x, \lambda} \leq 1 - 2 h^2(\h_1, \h_2)$. The convex hull of
these points is readily seen to be completely characterized by
\eqref{eq:dpeak} (together with the trivial inequalities $\abs{E_x} \leq 1$).

\begin{figure}[p]
  \centering
  \includegraphics{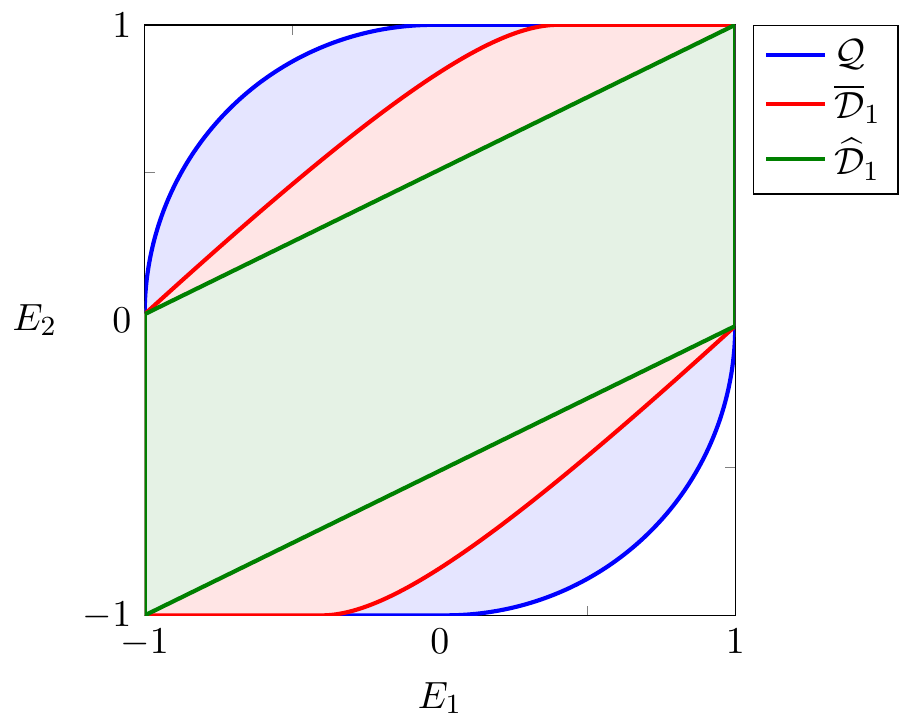}
  \caption{Comparison of the sets $\setDa_1$ and $\setDp_1$, corresponding
    respectively to the \emph{max-average} and \emph{max-peak} assumptions,
    for $\h_1 = \h_2 = 0.15$, illustrating that the latter of the two
    assumptions puts a stronger constraint on the behavior of the
    devices. The quantum set $\setQ$ is also represented for comparison.\label{fig:peak}}
\end{figure}

The sets $\setDa_1$ and $\setDp_1$ are compared in
Fig.~\ref{fig:peak}. Note that when $\h_1 = 0$ or $\h_2 = 0$,
$\setDa_1 = \setDa_2 = \setCa$.

\subsection{Implementation examples}
\label{sec:impld}

\begin{figure}[p]
  \centering
  \begin{subfigure}{0.45\linewidth}
  \centering
    \includegraphics{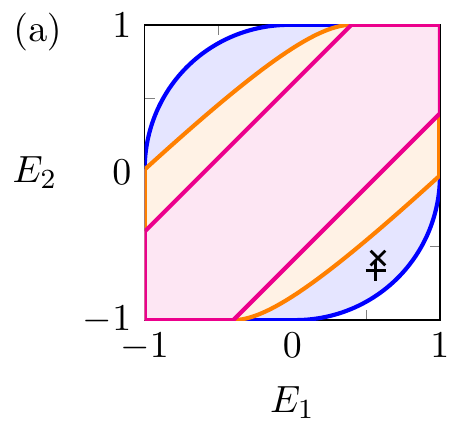}
    \phantomcaption{\label{fig:points_avg_eq}}
  \end{subfigure}%
  \begin{subfigure}{0.55\linewidth}
    \includegraphics{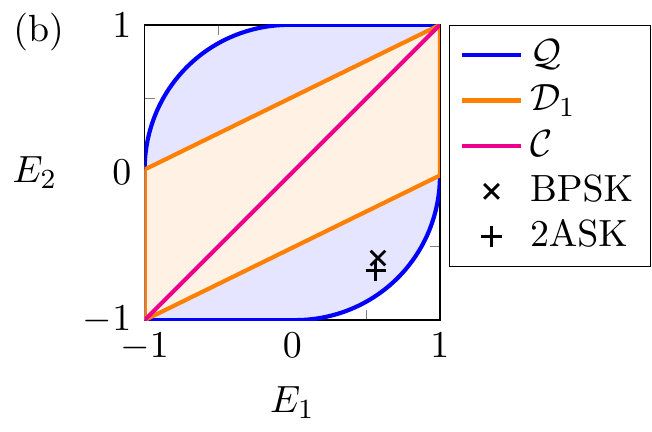}
    \phantomcaption{\label{fig:points_pk_eq}}
  \end{subfigure}
	
  \begin{subfigure}{0.45\linewidth}
  \centering
    \includegraphics{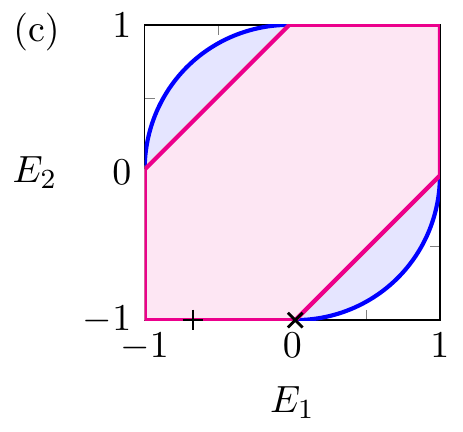}
    \phantomcaption{\label{fig:points_avg_vac}}
  \end{subfigure}%
  \begin{subfigure}{0.55\linewidth}
    \includegraphics{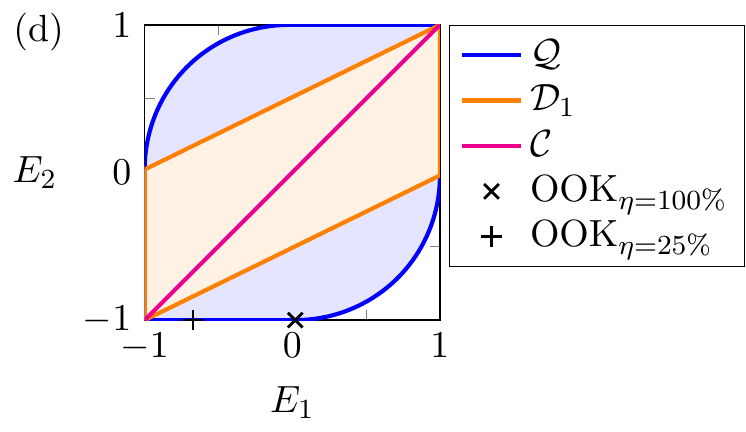}%
    \phantomcaption{\label{fig:points_pk_vac}}
  \end{subfigure}
  \caption{The sets $\mathcal{Q}$, $\mathcal{C}$, and $\mathcal{D}_1$ are
    displayed and compared to the correlations produced with the three
    implementations. The BPSK and the 2ASK protocols are analyzed with the
    constraints $\h_{1} = \h_{2} = 0.15$ under \subref{fig:points_avg_eq} the
    average-peak assumption and \subref{fig:points_pk_eq} the max-peak
    assumption. The OOK protocol is analyzed for two different detector
    efficiencies ($\eta = 100\%$ and $\eta = 25\%$) with the constraints
    $\h_1 = 0.51$ and $\h_2 = 0$ under \subref{fig:points_avg_vac} the
    average-peak assumption and \subref{fig:points_pk_vac} the max-peak
    assumption. Note that when $\h_{2} = 0$, the sets $\setCa$ and $\setDa_1$
    coincide under the max-average assumption.\label{fig:points}}
\end{figure}

The three experimental implementations that we presented in
Subsection~\ref{sec:impl} can be used to certify the production of genuine
randomness, as illustrated in Fig.~\ref{fig:points}. The BPSK and 2ASK
implementations can generate certified randomness under the max-average
assumption, while the OOK implementation requires the max-peak assumption.

\begin{figure}[p]
  \centering
  \includegraphics{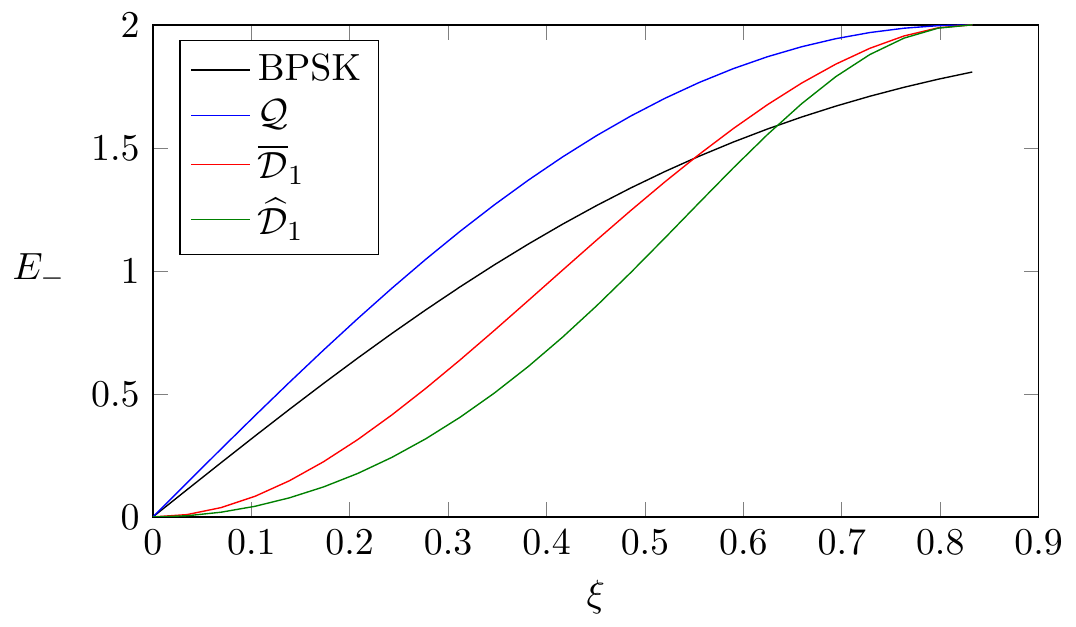}
  \caption{The correlations produced with the BPSK implementation are
    compared to the sets $\setDa_1$ and $\setDp_1$. As these correlations lie
    on the $E_-$-axis (they satisfy $E_1 + E_2 = 0$), it is only necessary to
    compare the value of $E_-$ for the BPSK implementation to the
    intersections of the boundaries of the sets $\setDa_1$ and $\setDp_1$
    with the $E_1$-axis. This is done for different values of the parameter
    $\xi$, which determines corresponding values for $\h_{1}$ and $\h_{2}$
    through \eqref{eq:thresh1}. The range
    $0 \leq \xi \leq \sqrt{\ln 2} \approx 0.83$ was chosen such that
    $0 \leq \h_1, \h_2 \leq 0.5$. This figure shows that, with the BPSK
    implementation, it is possible to generate correlations that produce
    certifiable randomness for $\xi \lesssim 0.55$ under the max-average
    assumption and for $\xi \lesssim 0.63$ under the max-peak assumption.}
  \label{fig:bpsk_parameters}
\end{figure}

Note that the BPSK correlations \eqref{eq:corr1} satisfy
$E_+ = E_1 + E_2 = 0$ and thus only the values of $\abs{E_-}$ are important
to determine whether they are outside of $\setDa_1$ or
$\setDp_1$. Fig.~\ref{fig:bpsk_parameters} compares the BPSK value of
$\abs{E_-}$ to the intersections of the sets $\setDa_1$ and $\setDp_1$ with
the $E_-$-axis as a function of the parameter $\xi$.

Finally, it is clear from Fig.~\ref{fig:points} that the certification of
randomness is robust to noise in the three implementations, i.e., to
correlations that deviate from the ideal ones. Let us consider as an example
the OOK implementation in the case where the source emits a coherent state
$\ket{\xi}$ and the photodetector has a limited efficiency $\eta < 1$. The
correlations \eqref{eq:corr3} then change to
\begin{IEEEeqnarray}{c+c}
  \label{eq:corr3_eta}
  E_1 = 1 - 2 e^{-\abs{\xi}^2 \eta} \simeq -1 + 2 \abs{\xi}^2 \eta \,, &
  E_2 = -1 \,.
\end{IEEEeqnarray}
The inequality \eqref{eq:dpeak} characterizing the region $\setDp_1$ is
violated if $h^2(\h_1, \h_2) E_1 - E_2 > 1 - h^2(\h_1, \h_2)$. Inserting the
above values for $E_1$ and $E_2$ and the threshold values \eqref{eq:thresh3}
characterizing the source (which give $h(\h_1, \h_2) = e^{-\abs{\xi}^2/2}$), we
find the condition
\begin{equation}
  \bigro{1 - 2 e^{-|\xi|^2 \eta}} e^{-|\xi|^2} + 1 > 1 - e^{-|\xi|^2} \,,
\end{equation}
which is satisfied provided that $e^{-\abs{\xi}^2 \eta} < 1$, i.e., that
$\eta > 0$. In other words, the OOK implementation can generate certified
randomness with arbitrarily low detection efficiency in the absence of other
imperfections. The situation corresponding to $\eta=25\%$ is represented in
Fig.~\ref{fig:points_avg_vac} and Fig.~\ref{fig:points_pk_vac}. Since in
addition to this tolerance to detector inefficiency the OOK implementation is
also very simple to implement experimentally, we will present in a
forthcoming publication \cite{ref:vhPREP} a full theoretical analysis of a
semi-DI RNG protocol based on this scheme.

\section{Conclusion}
\label{sec:conclusion}

In this paper, we introduced a new setting for semi-device-independent
(semi-DI) quantum information. Contrarily to the usual approach, we do not
assume bounds on the Hilbert space dimension of the carriers of quantum
information, but instead on the mean values of one (or several) physical
observable(s). Ideally, the choice of such an observable should be dictated
by the physics of the source of quantum information carriers and rely as much
as possible on a high-level characterization of its internal behavior. In
quantum optics implementations, a natural choice is to upper bound the
expected number of photons of the states emitted by the source or, alternatively, the energy contained in a range of frequency modes describing the system. We have completely characterized analytically the set
of possible correlations in the simplest possible prepare-and-measure
scenario compatible with such an assumption. We have in particular identified
analogues of Bell inequalities, which are able to distinguish genuinely
quantum devices from those behaving in a purely classically pre-determined
fashion.

We note that semi-DI prepare-and-measure scenarios that do not rely on a
Hilbert space dimension bound have also been introduced in
\cite{ref:cbbb2015}. However, they rely on a bound on the average von Neumann
entropy of the emitted states, a quantity which as defined in \cite{ref:cbbb2015}
requires a greater level of characterization of the source and also
depends on the probability distribution $p(x)$ used to select the
preparation $x$.

Our approach has several interests. First of all, as with semi-DI approaches
based on dimension bounds, it is of the prepare-and-measure type, and thus
does not require the manipulation of entanglement. But, as with full DI
approaches based on non-locality, it relies on assumptions that are
physically motivated. It thus combines the practical advantages of these two
different approaches.

Quite nicely, such an advantage from the implementation point of view does
not come at the expense of theoretical simplicity. On the contrary, the
minimal requirements on the number of inputs and outputs in our setting
($x \in \{1, 2\}$, $y \in \{1\}$, $b \in \{\pm 1\}$) are smaller than those
required for non-locality or dimension-bound protocols. Furthermore, in this
minimal scenario, our assumptions are not only more natural than dimension
bounds, but also more restrictive since they force the emitted states to have
a sub-unit communication capacity.

A further point to notice is that the no-communication assumption in the
non-locality scenario and the dimension-bound assumption are ``yes or no''
criteria (though it is also possible to introduce more refined assumptions in
such contexts \cite{ref:spm2013,ref:wp2015}). Our assumptions are instead
formulated in term of parameters that can take a continuous range of possible
values, e.g., as thresholds on the average photon number. As such, they
naturally allow for the introduction of additional ``margins of security''
making protocols based on them more robust to device imperfections. For
instance, in an implementation using a source designed to prepare states with
certain average photon numbers, one could performing the analysis assuming
thresholds $\vect{\h}$ corresponding to higher average photon numbers in
order to allow for an additional margin for safety.

There are many potential applications of our results. First of all, note that
one could reverse their interpretation. We have characterized the possible
correlations $\vect{E}$ generated in a prepare-and-measure setting given
upper bounds on the expectation values $\vect{H}$ of a physical observable
$H$. But one can equally well understand these results as providing lower
bounds for $\vect{H}$ given that some correlations $\vect{E}$ are
observed. That is, in analogy to the concept of DI ``dimension witnesses''
\cite{ref:bp2008,ref:gb2010}, our results imply the existence, e.g., of DI
``photon-number witnesses''. It is also reasonable to expect that our results
could be exploited to perform self-testing of quantum properties, namely one
could probably infer that the states and measurements have a specific form if
certain correlations $\vect{E}$ are observed under our assumptions.

Our main motivation at the origin of the present paper, however, is the
possibility to introduce new, physically motivated semi-DI random number
generation and quantum key distribution protocols. We have seen in
Section~\ref{sec:impl} that very simple implementations of our
prepare-and-measure scenario can lead to correlations that do not admit any
deterministic explanation. It is in fact possible to use the characterization
of the quantum set that we have obtained here to compute precise lower bounds
on the randomness that is produced by such implementations as a function of
the correlations $\vect{E}$ they generate. Such lower bounds can directly be
combined with the analysis of \cite{ref:pam2010,ref:pm2013,ref:nbsp2016} and
then lead to explicit protocols for semi-DI random number generation
protocols. We will present in detail how to compute such lower bounds on the
randomness and analyze the resulting semi-DI RNG protocols under realistic
experimental condition in a forthcoming publication \cite{ref:vhPREP}, with a
special focus on the OOK implementation.

\paragraph{Note added} The use of the On-Off Keying protocol of
Fig.~\ref{fig:impl_ook} in the context of semi-DI randomness generation has
also been discussed by the authors of \cite{ref:bme2016}, although it was
analyzed under different technical and security assumptions.

\section*{Acknowledgments}

This work is supported by the Fondation Wiener-Anspach, the Interuniversity
Attraction Poles program of the Belgian Science Policy Office under the grant
IAP P7-35 photonics@be, the Spanish MINECO (Severo Ochoa grant SEV-2015-0522
and QIBEQI FIS2016-80773-P), the Generalitat de Catalunya (SGR 875 and CERCA
Program), the Fundaci\'o{} Privada Cellex, the AXA Chair in Quantum
Information Science, and the EU project QITBOX, S.P. and R.G.-P. are Research
Associates of the Fonds de la Recherche Scientifique
(F.R.S.-FNRS). N.J.C. acknowledges financial support from the Fonds de la
Recherche Scientifique (F.R.S.-FNRS) under grant T.0199.13.

%

\end{document}

%% file: jnls.tex
\def\jphysa{J.\ Phys.\ A: Math.\ Theor.}%

\def\jsovmath{J.\ Sov.\ Math.}
\def\nature{Nature}%
\def\natphys{Nature Phys.}%
\def\njp{New J.\ Phys.}%
\def\pra{Phys.\ Rev.\ A}%
\def\prl{Phys.\ Rev.\ Lett.}%
\def\prx{Phys.\ Rev.\ X}%
\def\rmp{Rev.\ Mod.\ Phys.}%